\documentclass{article}
\usepackage{amsmath}
\usepackage{graphicx}
\usepackage{amssymb}
\usepackage{subfigure}
\usepackage{cancel}
\usepackage{cite}
\usepackage{relsize}
\usepackage{bbold}
\usepackage[top=1in, bottom=1in, left=0.8in, right=0.8in]{geometry}
\usepackage{lineno}
\usepackage{tablefootnote}

\newcommand{\beq}{\begin{equation}}
\newcommand{\eeq}{\end{equation}}
\newcommand{\bea}{\begin{eqnarray}}
\newcommand{\eea}{\end{eqnarray}}
\newcommand{\bi}{\begin{itemize}}
\newcommand{\ei}{\end{itemize}}

\def\babar{\mbox{\slshape B\kern-0.1em{\smaller A}\kern-0.1em 
  B\kern-0.1em{\smaller A\kern-0.2em R}}}

\usepackage{color}
\usepackage{hyperref} 
\hypersetup{linktocpage=true}
\usepackage[all]{hypcap}
\hypersetup{
    bookmarks=true,         % show bookmarks bar?
    unicode=false,          % non-Latin characters in AcrobatÕs bookmarks
    pdftoolbar=true,        % show AcrobatÕs toolbar?
    pdfmenubar=true,        % show AcrobatÕs menu?
    pdffitwindow=false,     % window fit to page when opened
    pdfstartview={FitH},    % fits the width of the page to the window
    pdftitle={Manifestations of Warped Extra Dimension in Rare Charm Decays and Asymmetries.},    % title
    pdfauthor={Ayan Paul, Alejandro de La Puente and Ikaros I. Bigi},     % author
    pdfsubject={hep-ph},   % subject of the document
    pdfcreator={TeXShop},   % creator of the document
    pdfproducer={MacTex, PDFLaTeX}, % producer of the document
    pdfkeywords={CP Violation} {Rare Charm Decays} {Models with Warped Extra Dimension}, % list of keywords
    pdfnewwindow=true,      % links in new window
    colorlinks=true,       % false: boxed links; true: colored links
    linkcolor=blue,          % color of internal links
    citecolor=magenta,        % color of links to bibliography
    filecolor=magenta,      % color of file links
    urlcolor=cyan           % color of external links
}

\usepackage{fancyhdr}
\begin{document}
\thispagestyle{empty}
\vspace*{-22mm}
\begin{flushright}
UND-HEP-12-BIG\hspace*{.08em}09\\
\end{flushright}
\vspace*{9mm}

\begin{center}
{\Large {\bf\boldmath
Manifestations of Warped Extra Dimension in Rare Charm Decays and Asymmetries. }}
\vspace*{10mm}

{\bf  Ayan Paul$^a$, Alejandro de La Puente$^b$ and Ikaros I. Bigi$^c$} \\
\vspace{4mm}
{\small
{\sl $^a$Theoretical Physics, INFN, Sezione di Roma,}
{\sl  I-00185 Roma, Italy}\\\vspace{1mm}
{\sl $^b$Theory Group, TRIUMF,}
{\sl Vancouver BC V6T 2A3, Canada}\\\vspace{1mm}
{\sl $^c$Department of Physics, University of Notre Dame du Lac,}
{\sl Notre Dame IN 46556, USA}\\\vspace{1mm}

{\sl email: \href{mailto:ayan.paul@roma1.infn.it}{ayan.paul@roma1.infn.it}, \href{mailto:adelapue@triumf.ca}{adelapue@triumf.ca} and \href{mailto:ibigi@nd.edu}{ibigi@nd.edu}} \\
}

\vspace*{10mm}

{\bf Abstract}\vspace*{-1.5mm}\\
\end{center}

We study the effects of models with a warped extra dimension on rare charm decays. While the new degrees of freedom in these models are bounded above a few TeV, they can leave signatures in rare charm decays with their tiny Standard Model background. We look at several channels with leptons in the final state along with several asymmetries of both kinematical and dynamical nature. CPT invariance is `usable' for analyzing the impacts of new dynamics in $D$ meson decays and in particular, rare decays of charm mesons.

%\newpage
\tableofcontents
%\newpage
%\modulolinenumbers[5] % Uncomment to specify line number modulo.
%\linenumbers % Uncomment to start line numbering.

%%%%%%%%%%%%%%%%%%%%%
\section{Introduction}
%%%%%%%%%%%%%%%%%%%%%

One puzzle in high energy physics is the hierarchy that exists between the scale of electroweak symmetry breaking (EWSB) and the Planck scale, $M_{Pl}$, at which the effects of gravity must be taken into account. This hierarchy problem has inspired a great number of theories beyond the Standard Model (SM); one attractive solution being to assume the existence of a warped extra dimension. This class of models are known as Randall-Sundrum (RS) models~\cite{Randall:1999ee}. They make use of the geometry of the extra dimension to provide a natural explanation for the hierarchy between the EWSB scale and $M_{Pl}$. The RS geometry is built from a slice of $AdS_{5}$ bounded by two branes, where SM particles are allowed to propagate in the bulk between the electroweak or `infrared' (IR) brane and the Planck or `ultraviolet' (UV) brane. In order to properly address the hierarchy problem, the Higgs is localized near or at the IR brane and the size of the extra dimension is adjusted to properly redshift the Planck scale towards the electroweak scale. In addition to providing an attractive solution to the hierarchy problem, different incarnations of the RS model have been implemented to address the fermion mass hierarchy~\cite{Gherghetta:2000qt,Grossman:1999ra,Huber:2000ie,Huber:2003tu}. This is possible by adjusting the localization of the fermion fields relative to the Higgs five dimensional background profile. The existence of a warped extra dimension and the ability of matter to propagate in the bulk leads to the existence of an infinite tower of Kaluza-Klein (KK) modes in the four dimensional effective theory. Furthermore, different localizations of KK modes in the extra dimension can lead to sizable contributions to flavour changing neutral currents (FCNC) well beyond limits given by the data. Variations of the RS scenario have been constructed in order to suppress these contributions and remain consistent within current experimental constraints~\cite{Agashe:2004cp,Moreau:2006np}. 

The existence of electroweak precision data (EWPD) has placed very stringent bounds for the masses of KK modes. In particular, they constrain the masses of the lowest KK mode excitations to $M_{KK}\ge 12$ TeV for a strictly IR localized Higgs field~\cite{Huber:2000fh} and $M_{KK}\ge7$ TeV for a Higgs field localized near the IR brane~\cite{Cabrer:2011mw}. The LHC is on its way to probing large mass scales, but direct detection of new resonances will likely be within the 1 TeV regime. This has motivated further research to extend the RS model in order to place KK excitations within the reach of the LHC. In particular, work by the authors of~\cite{Agashe:2003zs} make use of an extended gauge group in the slice of $AdS_{5}$ to protect electroweak (EW) observables from  large contributions beyond experimental limits. They conclude that it is possible to fit the EWPD with KK masses around 3 TeV. Furthermore, the authors in~\cite{Agashe:2007ki,Agashe:2008jb} have argued that discovering these modes would be possible at the LHC with center of mass energies of 14 TeV and 100 fb$^{-1}$ of integrated luminosities. Models that modify the $AdS_{5}$ metric near the IR brane have also been shown to decrease the bounds on KK modes~\cite{Cabrer:2010si,Cabrer:2011fb,Cabrer:2011vu,Cabrer:2011mw}. Unlike the RS model, these extra dimensional structures are able to lower the KK mass scale to 2-3 TeV in large areas of their parameter space. Within this class of models, impact of KK modes  involving the third generation of quarks was shown to be within the reach of the LHC at center of mass energy of 8 TeV and 10 fb$^{-1}$ of integrated luminosities~\cite{Carmona:2011ib,deBlas:2012qf}.

The strongest constraint by far on this class of models comes from the very well measured values of $\epsilon_K$. In a comprehensive work \cite{Csaki:2008zd} addressing this issue, it was shown that in a general anarchic picture where the Yukawa couplings and the bulk mass parameters determined by geometry are in general of $O(1)$, the lower bound to the KK mass can be as high as 21 TeV. This is quite expectable as the $Q_{LR}$ operator, which is highly suppressed in the SM, does not need to be so constrained in this class of models. This in turn leads to large contributions to $\Delta F=2$ observables and is almost dangerous for this model when constraints from $\epsilon_K$ are considered. Since this class of models is analysed within an anarchic framework, it is very important to make sure the applied parameter space is allowed by the constraints from $\epsilon_K$. In the section in which we discuss the parameter space we will also point out how this constraint is respected in addition to respecting constraints from EWPD and flavour observables in the beauty and strange sectors. 

After the LHC 2012 run, both ATLAS and CMS have set lower bounds on KK Gravitons. The strongest of these bounds come from the decay of the Gravitons to  dilepton final states $(e^+e^-,\mu^+\mu^-)$. ATLAS sets this bounds at 2.68 TeV~\cite{Aad:2014cka} and CMS at 2.39 TeV~\cite{Chatrchyan:2012oaa} for $k/\overline{M}_{PL}=0.1$ where k is the space-time curvature of the extra dimension as described later and $\overline{M}_{PL}=M_{PL}/\sqrt{8\pi}$ is the reduced Planck mass. This sets a lower bound of about $M_{KK}>1.6$ TeV from the ATLAS data. This is a less stringent bound than those discussed above and is below the $M_{KK}$ that we consider in this analysis.

In this work, we study the effects of KK modes to rare decays of $D$ mesons within a RS framework with custodial isospin protection. As explained above, this class of models leads to KK mode excitations which are currently being probed by the LHC or will be in the near future. The structure of this model is a lot more complex, with the existence of an extended fermion sector. Furthermore, the flavour structure of this class of models leads to FCNCs already at the tree level. This has interesting implications for CP violating observables as well as rare decays of $K$, $B$ and $D$ mesons. Signatures of this class of models along with the correlations generated by the same in $K$ and $B$ physics have been studied in great detail by~\cite{Albrecht:2009xr,Duling:2009sf,Blanke:2008yr,Blanke:2008zb,Burdman:2003nt,Blanke:2012tv,Bauer:2009cf,Chang:2006si,Agashe:2004ay,Burdman:2002gr}. They have extracted the flavour structure of these models as well as the fermion mass hierarchy which is consistent with current experimental bounds. Studies have also been done of the dependence of flavour dynamics on KK mass scales. We try to compliment the work on $K$ and $B$ mesons with our study of the rare decays of neutral $D$ mesons. 

Table \ref{tab:rare} summarizes the theoretical and experimental status of the rare decay channels that we will study in this article. 

\begin{table*}
\begin{center}
\begin{tabular}{| l c c c |}
\hline
&&&\\
 {\bf OBSERVABLE} & {\bf SM SD} & {\bf SM LD} & {\bf EXPERIMENT}\\ \hline
&&&\\
BR($D^0\to\gamma\gamma$)&$(3.6-8.1)\times 10^{-12\;\dagger}$&$(1-3)\times 10^{-8}$\cite{Fajfer:2001ad,Burdman:2001tf}&$<2.4\times 10^{-6}$\cite{Lees:2011qz}\\
&&&\\
BR($D^0\to\mu^+\mu^-$)&$6\times 10^{-19\;\dagger}$&$(2.7-8)\times 10^{-13}$&$<6.2(7.6)\times 10^{-9}$\cite{Aaij:2013cza}\\
&&&\\
BR($D\to X_u\nu\bar\nu$)&$10^{-15}-10^{-16\;\dagger}$&$-$&$-$\\
&&&\\
BR($D^\pm\to X_u l^+l^-$)&$3.7\times 10^{-9\; \dagger}$&$\sim\mathcal{O}(10^{-6})$&$<\mathcal{O}(10^{-7})$\footnote{LHCb has set upper limits on {\it exclusive} decay modes with two charged leptons in the final state \cite{Aaij:2013sua,Aaij:2013uoa}. It is notable that these upper limits are lower than the LD theory estimates.}\\
&&&\vspace{-0.12in}\\
$A^{c}_{\rm FB}$&$\sim2\times 10^{-6\; \dagger}$&$-$&$-$\\
&&&\vspace{-0.12in}\\
$A^{c}_{\rm CP}$&$\sim 3\times 10^{-4\; \dagger}$&$-$&$-$\\
&&&\vspace{-0.12in}\\
$A_{\rm FB}^{\rm CP}$&$\sim 3\times 10^{-5\; \dagger}$&$-$&$-$\\
&&&\\
\hline
\end{tabular}
\caption{SM short distance (SD) and long distance (LD) contributions to $D^0\to\gamma\gamma$,  $D^0\to\mu^+\mu^-$, $D\to X_u\nu\bar\nu$ and $D^\pm\to X_u l^+l^-$. Detailed calculations of the numbers marked with a ($^\dagger$) can be found in the references \cite{Paul:2010pq,Paul:2011ar,Bigi:2011em} and a summary in reference \cite{Paul:2013tra}.} 
\label{tab:rare}
\end{center}
\end{table*}

The structure of this paper is as follows: In Section \ref{sec:RSC}, we introduce and review the main features of the RS model with custodial isospin protection. In Section \ref{sec:flavWD}, we discuss the flavour structure of the model along with the parameter space that we shall examine for $\Delta C=1$ processes. We calculate the new contributions to the effective Hamiltonian for the following $\Delta C=1$ observables in the decays: $D^0\to\mu^{+}\mu^{-}$, $D\to X_{u}\nu\bar{\nu}$ and $D\to X_{u}l^{+}l^{-}$ and discuss the results in Section~\ref{sec:RareDecays}. In Section \ref{sec:Asym} we look at the new dynamics (ND) signature in several asymmetries in $D\to X_{u}l^{+}l^{-}$. We discuss the correlations between several flavour observables in Sections \ref{sec:CorrDD} and \ref{sec:CorrBD}. In Section \ref{sec:AnaParam}  we analyse and summarise our results. We comment on the effects of this class of models on $\Delta C =2$ observables in Section \ref{sec:DC2}. Finally, we summarize our findings in Section \ref{sec:Sum}.

%%%%%%%%%%%%%%%%%%%%%
\section{A Randall-Sundrum model with custodial isospin}
\label{sec:RSC}
%%%%%%%%%%%%%%%%%%%%%

In this section we review the RS model with custodial isospin protection~\cite{Agashe:2003zs}. We examine how EWSB proceeds and its implication for the spectrum, in particular how it leads to FCNC at tree level. Additionally, we review the larger fermion structure of this model and how it leads to fermion mass mixing beyond the CKM structure.

%%%%%%%%%%%%%%%%%%%%
\subsection{Gauge structure and electroweak symmetry breaking}
%%%%%%%%%%%%%%%%%%%%

The RS model with custodial isospin is constructed in a slice of $AdS_{5}$ with metric:
\begin{equation}
ds^{2}=e^{-2A(y)}\eta_{\mu\nu}dx^{\mu}dx^{\nu}-dy^{2}.
\end{equation}
The extra dimension is restricted to an interval defined by two end points: a UV brane located at $L=0$ and an IR brane at a distance $y=L$ along the extra dimension. In order to soften contributions to the $S$ and $T$ parameters all gauge and matter fields are added in the bulk~\cite{Gherghetta:2000qt,Chang:1999nh}. This framework also provides a mechanism to generate the hierarchy of fermion masses~\cite{Grossman:1999ra,Huber:2000ie,Huber:2003tu,Agashe:2004cp}.

The RS model with custodial isospin consists of a bulk gauge symmetry given by
\begin{equation}
SU(3)_{c}\times SU(2)_{L}\times SU(2)_{R}\times U(1)_{X}\times P_{LR}.
\end{equation}
The $P_{LR}$ discrete symmetry interchanges the two $SU(2)$ gauge groups and it is implemented to suppress nonuniversal contributions to the $Zb_{L}\bar{b}_{L}$ vertex~\cite{Agashe:2006at}.
 In the bulk, the Lagrangian for the gauge sector is given by
\begin{eqnarray}
\nonumber L_{gauge}&=&\sqrt{|g|}\left(\frac{1}{4}TrW_{MN}W^{MN}-\frac{1}{4}Tr\tilde{W}_{MN}\tilde{W}^{MN}-\frac{1}{4}\tilde{B}_{MN}\tilde{B}^{MN}-\frac{1}{4}TrG_{MN}G^{MN}\right),
\end{eqnarray}
where $g$ denotes the 5D metric, $W_{MN}$ is the field strength for the $SU(2)_{L}$ gauge group, $\tilde{W}_{MN}$ for $SU(2)_{R}$, and $\tilde{B}_{MN}$ and $G_{MN}$ for $U(1)_{X}$ and QCD gauge groups respectively. By assigning the appropriate boundary conditions, (UV,IR), one is able to recover the EW gauge symmetry in the UV brane:
\begin{eqnarray}
W^{1,2,3}_{\mu}(++),~~~\tilde{B}_{\mu}(++),\nonumber \\
\tilde{W}^{1,2}_{\mu}(-+),~~~\tilde{W}^{3}_{\mu}(++),
\end{eqnarray}
where $(+)/(-)$ denote Neumann/Dirichlet boundary conditions. Furthermore, breaking of the bulk symmetry in the UV brane down to $SU(2)_{L}\times U(1)_{Y}$ leads to mixing between $\tilde{B}_{\mu}$ and $\tilde{W}^{3}_{\mu}$:
\begin{eqnarray}
Z'_{\mu}&=&\tilde{W}^{3}_{\mu}\cos\phi-\tilde{B}_{\mu}\sin\phi, \nonumber \\
B_{\mu}&=&\tilde{W}^{3}_{\mu}\sin\phi+\tilde{B}_{\mu}\cos\phi,
\end{eqnarray}
where 
\begin{eqnarray}
\cos\phi&=&\frac{g_{R}}{\sqrt{g^{2}_{R}+g^{2}_{X}}}, \nonumber \\
\sin\phi&=&\frac{g_{X}}{\sqrt{g^{2}_{R}+g^{2}_{X}}}.
\end{eqnarray}
Because the discrete symmetry $P_{L,R}$ interchanges $SU(2)_{L}$ and $SU(2)_{R}$, we have set the couplings $g_{L}=g_{R}$. The above symmetry breaking pattern guarantees massless zero modes for the fields with (+,+) boundary conditions.

Electroweak symmetry breaking is achieved as in the RS model~\cite{Randall:1999ee}, introducing a scalar Higgs field localized on the IR brane. This field transforms as a bidoublet of the $SU(2)_{R}\times SU(2)_{L}$ gauge symmetry and it is given by:
\begin{equation}
\Sigma=\begin{pmatrix}
\frac{\pi^{+}}{\sqrt{2}} & -\frac{h^{0}-i\pi^{2}}{2} \\
\\
\frac{h^{0}+i\pi^{2}}{2} & \frac{\pi^{-}}{\sqrt{2}}
\end{pmatrix}
\end{equation}
The neutral component of the Higgs field acquires a vacuum expectation value which breaks the EW bulk symmetry, $SU(2)_{R}\times SU(2)_{L}$, down to its diagonal combination $SU(2)_{V}$. As a consequence, the model retains an unbroken custodial symmetry which prevents the $S$ and $T$ parameters from receiving corrections that are enhanced by the volume of the extra dimension, $L\equiv kr\pi\approx \log \left(\frac{M_{PL}}{M_{W}}\right)\approx 37$~\cite{Casagrande:2010si}.

After EWSB, the photon and gluon zero modes remain massless and modes of fields coupling to the Higgs, $W^{1,2,3}_{\mu}$, acquire a mass. The following mixing pattern arises between the neutral components of the gauge fields:
\begin{eqnarray}
Z_{\mu}&=&W^{3}_{\mu}\cos\psi-B_{\mu}\sin\psi, \nonumber \\
A_{\mu}&=&W^{3}_{\mu}\sin\psi+B_{\mu}\cos\psi.
\end{eqnarray} 
The mixing angle $\psi$ can be expressed in terms of the angle $\phi$ using the following relations:
\begin{eqnarray}
\cos\psi&=&\frac{1}{\sqrt{1+\sin^{2}\phi}}, \nonumber \\
\sin\psi&=&\frac{\cos\phi}{\sqrt{1+\sin^{2}{\phi}}}.
\end{eqnarray}
The effective four-dimensional theory is then derived through the usual Kaluza-Klein reduction which yields an infinite tower of replicas for each of the gauge fields in the theory. For gauge bosons these are given by
\begin{equation}
A_{\mu}(x,y)=\frac{1}{\sqrt{L}}\sum_{n=0}f^{n}_{A}(y)A^{n}_{\mu}(x).
\end{equation}
Restricting the study to the case $n=0$ and $n=1$ KK modes, the neutral gauge boson mix, the neutral gauge boson interactions with the Higgs field lead to mixing between $Z'$ and the $n=0$ and $n=1$ modes of $Z$, $Z^{(0)}$ and $Z^{(1)}$. This mixing is parametrized by a unitary matrix, $R_{Z}$, and the mass eigenstates can be ordered such that
\begin{equation}
(Z^{0},Z^{1},Z^{'})^{T}=R_{Z}(Z_{3},Z_{2},Z_{1})^{T}.
\end{equation}
The lighter state, $Z_{1}$ corresponds to the SM neutral gauge boson while $Z_{2}$ and $Z_{3}$ have masses given by $M_{\text{KK}}\approx 2.45 f$, where
\begin{equation}
f=k e^{-kL}.
\end{equation}

%%%%%%%%%%%%%%%%%%%%%%
\subsection{Quarks and leptons}
%%%%%%%%%%%%%%%%%%%%%%

In the RS framework, an explanation for the fermion mass hierarchy and mixing is provided when fermions are placed in the bulk~\cite{Gherghetta:2000qt,Huber:2003tu}. Since fermions are embedded in representations of the bulk symmetry, the custodial isospin extension of the RS model suppresses the coupling of the $Zb_{L}\bar{b}_{L}$ vertex by preserving the $P_{LR}$ symmetry of the model~\cite{Carena:2006bn}. A well studied route has been to embed fermions within representations of the group $O(4)$, which is isomorphic to $SU(2)_{L}\times SU(2)_{R}\times P_{LR}$~\cite{Agashe:2006at}. Within this framework, left-handed {\em up}- and {\em down-}type quarks are embedded within bi-doublet representations of $S(2)_{L}\times SU(2)_{R}$, with the right-handed {\em up}-type quarks transforming as singlets under $O(4)$. The right-handed {\em down}-type quarks transform as a $({\bf 3,1 })\oplus ({\bf 1,3 })$. As with gauge bosons, fields with Neumann boundary conditions at both branes, $(++)$, correspond to the SM model spectrum. The electric charges for fermions are given by
\begin{equation}
Q=T^{3}_{L}+T^{3}_{R}+Q_{X},
\end{equation} 
where $T^{3}_{L,R}$ denote the third component of $SU(2)_{L,R}$ and $Q_{X}$ the $U(1)_{X}$ charge. Through this framework, all fermions receive the same $U(1)_{X}$ charge and after electroweak symmetry breaking, mixing occurs between fermions with the same electric charge. A detailed study of the mixing between zero modes and the lowest KK modes is carried out in~\cite{Albrecht:2009xr}, where the structure of the model consists of three mass matrices for the states with charges $5/3$, $2/3$, and $-1/3$.

The bulk symmetry preserving are given by
\begin{eqnarray}\label{eq:4Dyuk}
\left(Y^{(5/3)}_{ij}\right)_{kl}&=&\frac{1}{N}\int^{L}_{0}dy\lambda^{d}_{ij}f^{(5/3)}_{L,k}(y)f^{(5/3)}_{R,l}(y)h(y), \nonumber \\
\left(Y^{(2/3)}_{ij}\right)_{kl}&=&\frac{1}{N}\int^{L}_{0}dy\lambda^{d}_{ij}f^{(2/3)}_{L,k}(y)f^{(2/3)}_{R,l}(y)h(y), \nonumber \\
\left(\tilde{Y}^{(2/3)}_{ij}\right)_{kl}&=&\frac{1}{N}\int^{L}_{0}dy\lambda^{u}_{ij}f^{(2/3)}_{L,k}(y)f^{(2/3)}_{R,l}(y)h(y), \nonumber \\
\left(Y^{(-1/3)}_{ij}\right)_{kl}&=&\frac{1}{N}\int^{L}_{0}dy\lambda^{d}_{ij}f^{(-1/3)}_{L,k}(y)f^{(-1/3)}_{R,l}(y)h(y),\nonumber\\
\end{eqnarray}
where $N = \sqrt{2}L^{3/2}$. The function $h(y)$ is the Higgs profile function given by
\begin{equation}
h(y)=\sqrt{2(\beta-1)kL}e^{kL}e^{\beta k(y-L)}
\end{equation}
with $\beta\gg 1$ for IR localized Higgs and the functions $f_{L,R}$ are the fermion 5D profiles given in~\cite{Gherghetta:2000qt,Huber:2000ie}. The fermionic zero mode is given by:
\begin{equation}
f^{0}_{L,R}\left(y\right)=\sqrt{\frac{(1\mp 2c)kL}{e^{(1\mp2c)kL}-1}}e^{\mp cky},
\end{equation}
where $c$ is the bulk mass parameter. The zero mode is localised near the IR brane for $c<1/2$ and near the UV brane for $c>1/2$. In the lepton sector, left-handed charged leptons and neutrinos are grouped within a bi-doublet representation of $SU(2)_{L}\times SU(2)_{R}$, and the right-handed charged leptons transform as a $({\bf 3,1 })\oplus ({\bf 1,3 })$ of the EW gauge symmetry. In addition, this model contains a right-handed neutrino field transforming as a singlet under the $O(4)\times U(1)_{X}$ bulk symmetry.

%%%%%%%%%%%%%%%%%%%%%%%%%%%%%%%%
\subsection{Couplings of fermions to neutral gauge bosons}
%%%%%%%%%%%%%%%%%%%%%%%%%%%%%%%%

To calculate the tree level contributions to FCNCs we first derive the couplings of fermions to the electroweak neutral gauge bosons. Because of the mixing that arises between the zero and first KK modes of $Z_{\mu}$ with $Z'_{\mu}$, we must diagonalize the mass terms in order to find the appropriate mass eigenstates. In this section we will work in the EW basis for fermions. The rotation to mass eigenstates will be shown in the following section, where the CKM structure of the model will be discussed in greater depth. In order to derive the couplings between neutral gauge bosons and {\em up}-type fermions we follow the analysis in~\cite{Agashe:2006at}. The left-handed {\em up}-type quarks transform as a bidoublet of the full $O(4)\times U(1)_{X}$ symmetry and couple to the gauge bosons through the following Lagrangian:
\begin{eqnarray}
{\cal L}\supset &&\; a_{1}Tr[\bar{Q}_{L}\gamma^{\mu}Q_{L}\hat{V}_{\mu}]+a_{2}Tr[\bar{Q}_{L}\gamma^{\mu}V_{\mu}Q_{L}]+a_{3}Tr[\bar{Q}_{L}\gamma^{\mu}iD_{\mu}U]Tr[U^{\dagger}Q_{L}]+h.c\label{eq:lageff},
\end{eqnarray}
where $Q_{L}\in ({\bf 2,2})_{2/3}$ of $O(4)\times U(1)_{X}$, $V_{\mu}=(iD_{\mu}U)U^{\dagger}$, and $\hat{V}_{\mu}=(iD_{\mu}U)^{\dagger}U$. We make use of the following covariant derivative
\begin{equation}
D_{\mu}U=\partial_{\mu} U+i\frac{g_{L}}{2}\sigma_{a}W^{a}_{\mu}U+i\frac{g_{R}}{2}\sigma_{a}\tilde{W}^{a}_{\mu}U-i\frac{g_{X}}{2}\tilde{B}_{\mu}U\sigma_{3},
\end{equation}
where $U$ is a non-linear sigma field which contains the pseudo Nambu-Goldstone bosons resulting from the following breaking pattern: $O(4)\to O(3)$. The $P_{LR}$ symmetry of the model imposes $g_{L}=g_{R}$ and $a_{1}=a_{2}$; guaranteeing the vanishing of large dangerous contributions to the $Zb_{L}\bar{b}_{L}$ coupling. In the language of extra dimensions, the custodial protection mechanism prevents the $Zb_{L}\bar{b}_{L}$ from corrections enhanced by the volume of the extra dimension~\cite{Casagrande:2010si}. 
 
We have naively used effective four dimensional fields together with the fundamental 5D gauge coupling. To obtain the effective 4D coupling, integration over the fermion and gauge boson profiles must be carried out with $U=\mathbb{1}_{2\times2}$ after the $O(4)$ symmetry is broken. 

The couplings of left-handed {\em up}-type quarks to the $Z^{0}$, $Z^{1}$ and $Z'$ neutral gauge bosons are given by
\begin{eqnarray}
g^{q}_{L,Z^{0}}&=&\frac{g_{4}}{\cos\psi}\left(\frac{1}{2}-\frac{2}{3}\sin^{2}\psi\right)F^{0}_{L}(c^{q}_{L,i},c^{q}_{L,i}), \nonumber \\
g^{q}_{L,Z^{1}}&=&\frac{g_{4}}{\cos\psi}\left(\frac{1}{2}-\frac{2}{3}\sin^{2}\psi\right)F^{1}_{L}(c^{q}_{L,i},c^{q}_{L,i}), \nonumber \\
g^{q}_{L,Z'}&=&\frac{g_{4}}{\cos\phi}\left(-\frac{1}{2}-\frac{1}{6}\sin^{2}\phi\right)F'_{L}(c^{q}_{L,i},c^{q}_{L,i}),\label{eq:leftg}
\end{eqnarray}
and those of right-handed {\em up}-type quarks by
\begin{eqnarray}
g^{q}_{R,Z^{0}}&=&-\frac{2}{3}\frac{g_{4}}{\cos\psi}\sin^{2}\psi F^{0}_{R}(c^{q}_{R,i},c^{q}_{R,i}), \nonumber \\
g^{q}_{R,Z^{1}}&=&-\frac{2}{3}\frac{g_{4}}{\cos\psi}\sin^{2}\psi F^{1}_{R}(c^{q}_{R,i},c^{q}_{R,i}), \nonumber \\
g^{q}_{R,Z'}&=&-\frac{2}{3}\frac{g_{4}}{\cos\phi}\sin^{2}\phi F'_{R}(c^{q}_{R,i},c^{q}_{R,i}),\label{eq:rightg}
\end{eqnarray}
where $g_{4}$ denotes the $SU(2)_{L}$ coupling strength as in the SM model, and $F^{0,1,}$$'_{L,R}(c_{L,R,i},c_{L,R,i})$ are entries of the diagonal matrices that quantify the overlap between right- and left-handed zero-mode fermions with the $Z'^{,0,1}$ neutral gauge bosons:
\begin{eqnarray}
(F^{0}_{L,R})_{ii}=\int e^{ky}f^{0,c_{i}}_{L,R}(y)f^{0,c_{i}}_{L,R}(y)f^{0}_Z(y)dy\nonumber\\
(F^{1}_{L,R})_{ii}=\int e^{ky}f^{0,c_{i}}_{L,R}(y)f^{0,c_{i}}_{L,R}(y)f^{1}_Z(y)dy\nonumber \\
(F'_{L,R})_{ii}=\int e^{ky}f^{0,c_{i}}_{L,R}(y)f^{0,c_{i}}_{L,R}(y)f^{1}_{Z'}(y)dy
\end{eqnarray}

The index $i=1,2,3$, denotes the fermion family number, $f_Z^{n}(y)$ the gauge boson profile of the  $n^{\text{th}}$ KK mode of the $Z$ boson and $f_{Z'}^{n}(y)$ that for the $Z'$ gauge boson. The coupling to the photon is given by
\begin{equation}
g^{q}_{L,R}=\frac{2}{3}g_{4}\sin\psi.\label{eq:photong}
\end{equation}

The analysis of the interactions between neutral gauge bosons and leptons is carried out in the same way. The interactions of the left-handed leptons are obtained from Equation~(\ref{eq:lageff}), and lead to the following effective couplings:
\begin{eqnarray}
g^{l}_{L,Z^{0}}&=&-\frac{g_{4}}{\cos\psi}\left(\frac{1}{2}\sin^{2}\psi\right)F^{0}_{L}(c^{l}_{L,i},c^{l}_{L,i}), \nonumber \\
g^{l}_{L,Z^{1}}&=&-\frac{g_{4}}{\cos\psi}\left(\frac{1}{2}\sin^{2}\psi\right)F^{1}_{L}(c^{l}_{L,i},c^{l}_{L,i}), \nonumber \\
g^{l}_{L,Z'}&=&\frac{g_{4}}{\cos\phi}\left(-\frac{1}{2}+\frac{1}{2}\sin^{2}\phi\right)F'_{L}(c^{l}_{L,i},c^{l}_{L,i}).\label{eq:gLl}
\end{eqnarray}
Right-handed leptons belong to the ${\bf (1,3)}$ triplet representation of $SU(2)_{L}\times SU(2)_{R}$ and an effective Lagrangian analysis is used to derive their couplings to neutral gauge bosons~\cite{Agashe:2006at}. These can be parametrized by the following effective Lagrangian 
\begin{equation}
{\cal L}\supset a_{4}Tr[\bar{E}_{R}\gamma^{\mu}E_{R}\hat{V}_{\mu}]+a_{5}Tr[\bar{E}_{R}\gamma^{\mu}V_{\mu}E_{R}].
\end{equation}
The effective couplings are given by:
\begin{eqnarray}
g^{l}_{R,Z^{0}}&=&\frac{g_{4}}{\cos\psi}\sin^{2}\psi F^{0}_{R}(c^{l}_{R,i},c^{l}_{R,i}), \nonumber \\
g^{l}_{R,Z^{1}}&=&\frac{g_{4}}{\cos\psi}\sin^{2}\psi F^{1}_{R}(c^{l}_{R,i},c^{l}_{R,i}), \nonumber \\
g^{l}_{R,Z'}&=&-g_{4}\cos\phi F'_{R}(c^{l}_{R,i},c^{l}_{R,i}),\label{eq:gRl}
\end{eqnarray}
while those to the photon KK mode are given by $g^{l}_{(L,R),\gamma}=-g_{4}\sin\psi$.

In Section~\ref{sec:RareDecays} we show the new tree-level contributions to the rare decays of $D$ mesons due to the couplings introduced above using a KK mass scale of $2.45$ TeV. Before proceeding, in Section~\ref{sec:flavWD} we discuss the flavour structure of the model and the parameter sets used in the analysis of $D$ meson rare decays.

%%%%%%%%%%%%
\section{The flavour structure and the parameter sets}
\label{sec:flavWD}
%%%%%%%%%%%%

\begin{figure*}[t]
\includegraphics[width=17cm]{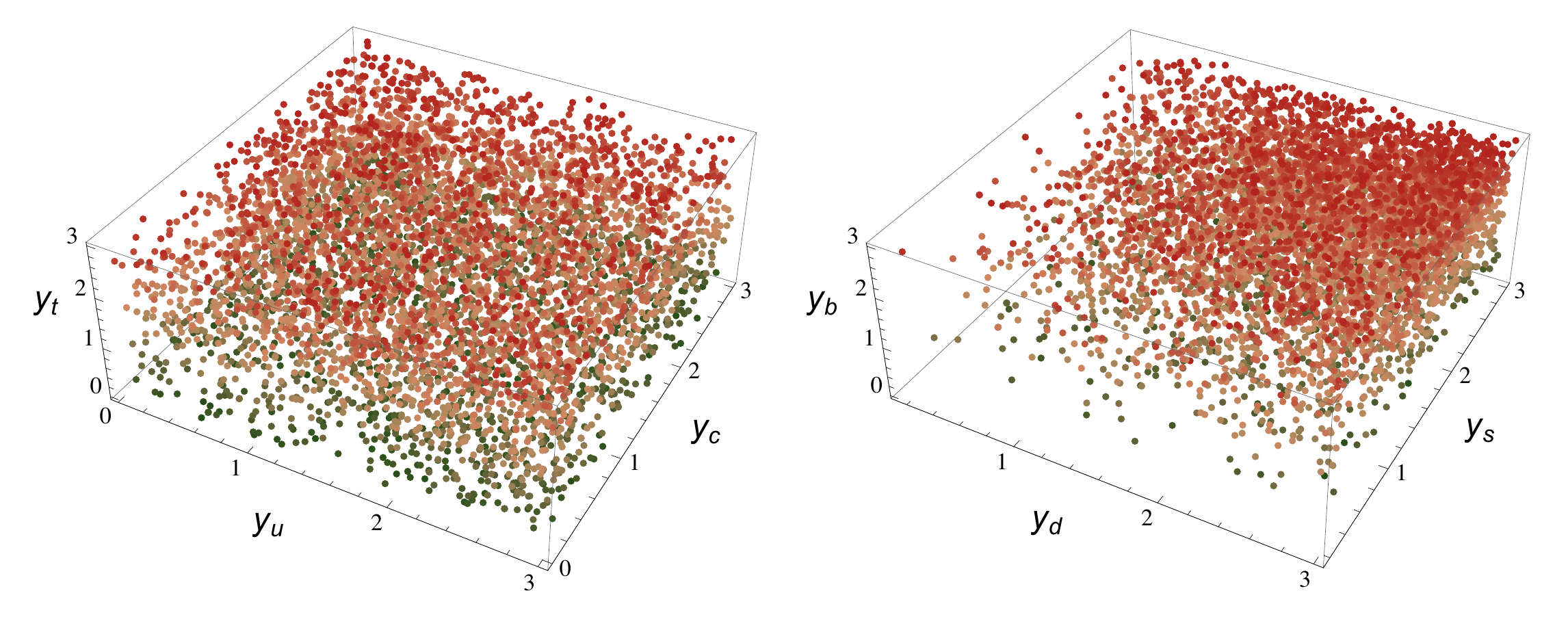}
\caption{Scatter plot of the Yukawa couplings of the quarks. The figure on the left is for the {\em up-}type quarks and that on the right is for the {\em down-}type quarks.}
\label{fig:Yukawa}
\end{figure*}

The flavour structure of this model was studied in depth in~\cite{Blanke:2008zb,Albrecht:2009xr}. Alternative approaches can be found in \cite{Casagrande:2008hr,Bauer:2009cf}. Before looking at the structure of the CKM matrix, it is worth emphasizing that the mixing of fermionic zero modes with heavier fermionic KK modes, also leads to a source of FCNC. However as it was pointed out in~\cite{Blanke:2008zb} as well as~\cite{Agashe:2004cp,Casagrande:2008hr}, these effects are negligible and thus we only consider the $k=l=0$ components of the 4D Yukawa matrices, Equation~(\ref{eq:4Dyuk}), when rotating to the mass eigenstate basis. The 4D Yukawas can then be effectively written as
\begin{equation} 
Y^{u,d}_{ij}\equiv\lambda^{u,d}_{ij}\frac{e^{kL}}{kL}f^{0}_{L}(y=L,c^{i}_{Q})f^{0}_{R}(y=L,c^{j}_{u,d}),
\end{equation}
where $\lambda^{u,d}_{ij}$ denote the 5D Yukawa couplings. The CKM matrix is obtained as in the SM, that is
\begin{equation}
V_{CKM}=U^{\dagger}_{L}D^{}_{L},
\end{equation}
where $U_{L}$ and $D_{L}$ are unitary matrices which rotate flavour eigenstates into mass eigenstates for left-handed up- and {\em down}-type quarks. Their complete parametrization can be found in~\cite{Blanke:2008zb,Albrecht:2009xr}. Since we are considering neutral current exchanges at tree level, we found it useful to define effective rotation matrices for the {\em up}-type quarks, $U^{q}_{L,eff}$ and $U^{q}_{R,eff}$ coupling quark mass eigenstates to the three physical massive neutral gauge bosons as well as to the first KK mode of the photon. These can be obtained using Equations~(\ref{eq:leftg}),(\ref{eq:rightg}), and~(\ref{eq:photong}):
\begin{eqnarray}
U^{q}_{L,eff}(n)&=&g_{L,Z^{0}}\cdot R_{Z,1n}+U_{L}^{\dagger}g^{q}_{L,Z^{1}}U_{L}\cdot R_{Z,2n}+U_{L}^{\dagger}g^{q}_{L,Z'}U_{L}\cdot R_{Z,3n}, \nonumber \\
U^{q}_{R,eff}(n)&=&g^{q}_{R,Z^{0}}\cdot R_{Z,1n}+U_{R}^{\dagger}g^{q}_{R,Z^{1}}U_{R}\cdot R_{Z,2n}+U_{R}^{\dagger}g^{q}_{R,Z'}U_{R}\cdot R_{Z,3n},\label{eq:UqZ}
\end{eqnarray}
for the three physical neutral gauge bosons denoted by $n=1,2,3$ and
\begin{eqnarray}
U^{q}_{L,eff}(\gamma ')&=&U_{L}^{\dagger}g^{q}_{L,\gamma}U_{L}, \nonumber \\
U^{q}_{R,eff}(\gamma ')&=&U_{R}^{\dagger}g^{q}_{R,\gamma}U_{R}, \label{eq:UqA}
\end{eqnarray}
for the first KK mode of the photon. In a similar way, for leptons, we use Equations~(\ref{eq:gLl}) and~(\ref{eq:gRl})
\begin{eqnarray}
U^{l}_{L,eff}(n)&=&g^{l}_{L,Z^{0}}\cdot R_{Z,1n}+g^{l}_{L,Z^{1}}\cdot R_{Z,2n}+g^{l}_{L,Z'}\cdot R_{Z,3n}, \nonumber \\
U^{l}_{R,eff}(n)&=&g^{l}_{R,Z^{0}}\cdot R_{Z,1n}+g^{l}_{R,Z^{1}}\cdot R_{Z,2n}+g^{l}_{R,Z'}\cdot R_{Z,3n},\label{eq:UlZ}\nonumber\\
\end{eqnarray}
for the three physical neutral gauge bosons denoted by $n=1,2,3$ and
\begin{eqnarray}
U^{l}_{L,eff}(\gamma ')&=&g^{l}_{L,\gamma}, \nonumber \\
U^{l}_{R,eff}(\gamma ')&=&g^{l}_{R,\gamma}, \label{eq:UlA}
\end{eqnarray}
for the first KK mode of the photon.

In what follows we describe how the parameters encoded in the matrices $g^{q,l}_{L,R}$ are constrained and how fine tuned they need to be to follow these constraints.

Being motivated by ``naturalness'' and an attempt to generate a natural hierarchy from the Plank scale down to the electroweak scale, the models with a warped extra dimension are at their best when the absence of fine tuning is prevalent in all parameters. However, severe constraints from flavour dynamics, especially from the measured value of $\epsilon_K$ force some degree of fine tuning in these models. Even if these models have a custodial symmetry that protects them from electroweak precision constraints, further constraints on the KK mass scale are generated due to constraints from $\epsilon_K$. However, all of this can be alleviated by a judicious choice of the parameter space that we deem as allowed. A ``judicious'' choice always means the reintroduction of tuning of the parameter sets. However, what is important here is not the presence or absence of tuning, but the degree to which this is necessary. The primary aim of this is to keep the KK mass scale as low as possible to make the model testable in the near future, while not allowing the violation of flavour constraints. 

The parameter set used in this analysis was generated and used previously by the authors of \cite{Blanke:2008zb,Blanke:2008yr} to study $B$ and $K$ physics. A detailed explanation of the generation of the parameter set and its check against constraints have been discussed elaborately in those works. We cannot do it any more justice. However, we will address a few concerns that are commonly raised when such models are discussed.

\begin{itemize}
\item The parameter set used in this work was generated with a breaking scale of $f=1$ TeV which gives a KK mass scale of $2.45$ TeV. This seriously brings into question the feasibility of such low KK scales in the light of the analysis done in \cite{Csaki:2008zd}. It should be pointed out here that these constraints apply to the RS model with anarchic Yukawa and bulk fermion mass parameters in general, but does not rule out the possibility of ``psedo-anarchy'' in parts of the parameter space. 
\item Such ``peudo-anarchy'', or possible fine tuning can be aptly parametrized by the measure proposed by Barbieri and Giudice \cite{Barbieri:1987fn} $\Delta_{BG}(O_i,p_i)$. The authors of \cite{Blanke:2008zb,Blanke:2008yr} took this into consideration when scanning the parameter space for points allowed by the constraints discussed previously. In particular they allowed only points in the parameter space which satisfied $\Delta_{BG}(\epsilon_K)<20$.
\item It was also shown in \cite{Blanke:2008zb,Blanke:2008yr}, that although the average fine tuning necessary for the accommodation of the constraint from $\epsilon_K$, there lies a part of the parameter space that is not excluded once moderate fine-tuning, as defined above, is allowed. 
\end{itemize}

In the light of this argument we shall display the parameter set that has been used in the analysis in a manner that lays it open to naturalness arguments. Amongst the parameters relevant to this analysis only a subset yield to this argument on naturalness, namely, the Yukawa couplings of the quark sector. We do not discuss the naturalness in the Yukawa couplings of the lepton sector as it is not relevant to this analysis.\footnote{In this analysis all three families of leptons are all localised with the same bulk mass parameter which also allows one the added benefit of escaping constraints from lepton FCNC.}

The Yukawa coupling of the {\em up-}type quarks seem to be quite ``anarchic'' and random as can be seen from Figure~\ref{fig:Yukawa}, being evenly distributed over the entire parameter space. On the other hand, the {\em down-}type Yukawa couplings seem to have a tendency to cluster towards the higher values of the couplings for the first and the second family. It must be kept in mind that the allowed parameter set is shaped by both constraints from electroweak precision tests and flavour constraints, some of which are stronger on the first two families than on the third, especially for the {\em down}-type quarks.

%%%%%%%%%%%%
\section{Rare decays in charm dynamics}
\label{sec:RareDecays}
%%%%%%%%%%%%

FCNC from this model of ND comes from tree level exchanges of neutral gauge bosons. The primary contribution on all the decay channels that we have studied, and elucidated on below, comes from the mixing of the new gauge boson states, into $Z_{1}$, the SM neutral massive gauge boson. Contributions that arise from $Z_{2}$, $Z_{3}$ and $A^{(1)}$ are subdominant or negligible. However, we keep their contributions in the formulation for completion. It also serves the purpose of showing that the addition of the higher KK modes of the neutral gauge bosons can only produce infinitesimal contributions. Taking into consideration the nature of the dependence of the contributions on the mass of the gauge states, these contributions will steadily decrease in magnitude as we move up the tower and hence fail to introduce considerable enhancements. The corrections from the higher fermion modes in the KK tower can also be similarly argued to be subleading.

%%%%%%%%%%%%
\boldmath
\subsection{$D^0\to\mu^+\mu^-$}
\unboldmath
%%%%%%%%%%%%

The SM LD contribution to $D^0\to \mu^+\mu^-$ is driven by the total branching fraction of $D^0\to\gamma\gamma$ \cite{Burdman:2001tf,Fajfer:2001ad} and is independent of whether the latter is generated by SM or by ND \cite{Paul:2010pq}. Hence, enhancements to SM LD contributions to $D^0\to\mu^+\mu^-$ can be generated by ND contributions to the $D^0\to\gamma\gamma$ channel. Using the notations from \cite{Paul:2010pq}, the contributions of this model can arise in the 1PR  and the 2PR (or 1PI) contributions. ND intervenes only at the loop level and both from the same sources.

\begin{itemize}
\item From the mixing of $W^{1\pm}$ and $W'^\pm$ with the $W^{0\pm}$ which gives the physical $W^{\pm}_{1}$ states.
\item From the heavier physical states $W^\pm_{2}$ and $W^\pm_{3}$ driving charged currents with SM fermions.
\end{itemize}

The possibilities of large contribution of ND from the warped extra dimension in $\Delta C=1$ processes are solely because FCNCs are generated at the tree level itself. However, for this channel, due to the structure of the contributing diagrams, it is not possible for tree level ND to enhance the rate. The leading contributions come only at the loop level as additional gauge states in the loop for the 2PR diagram and enhancements to the effective $\gamma$ vertex for the 1PR diagram. Hence, SM LD contributions will dominate over any new contributions that ND can generate considering the former is almost three orders of magnitude larger than the SM SD contributions (cf. Table \ref{tab:rare}). This leads us to the conclusion that the SM LD rate for $D^0 \to \mu^+\mu^-$ remains controlled by purely SM contributions to $D^0\to \gamma \gamma$ even in the presence of ND from the warped extra dimension.

\begin{figure}[h!]
\begin{center}
\includegraphics[width=8.35cm]{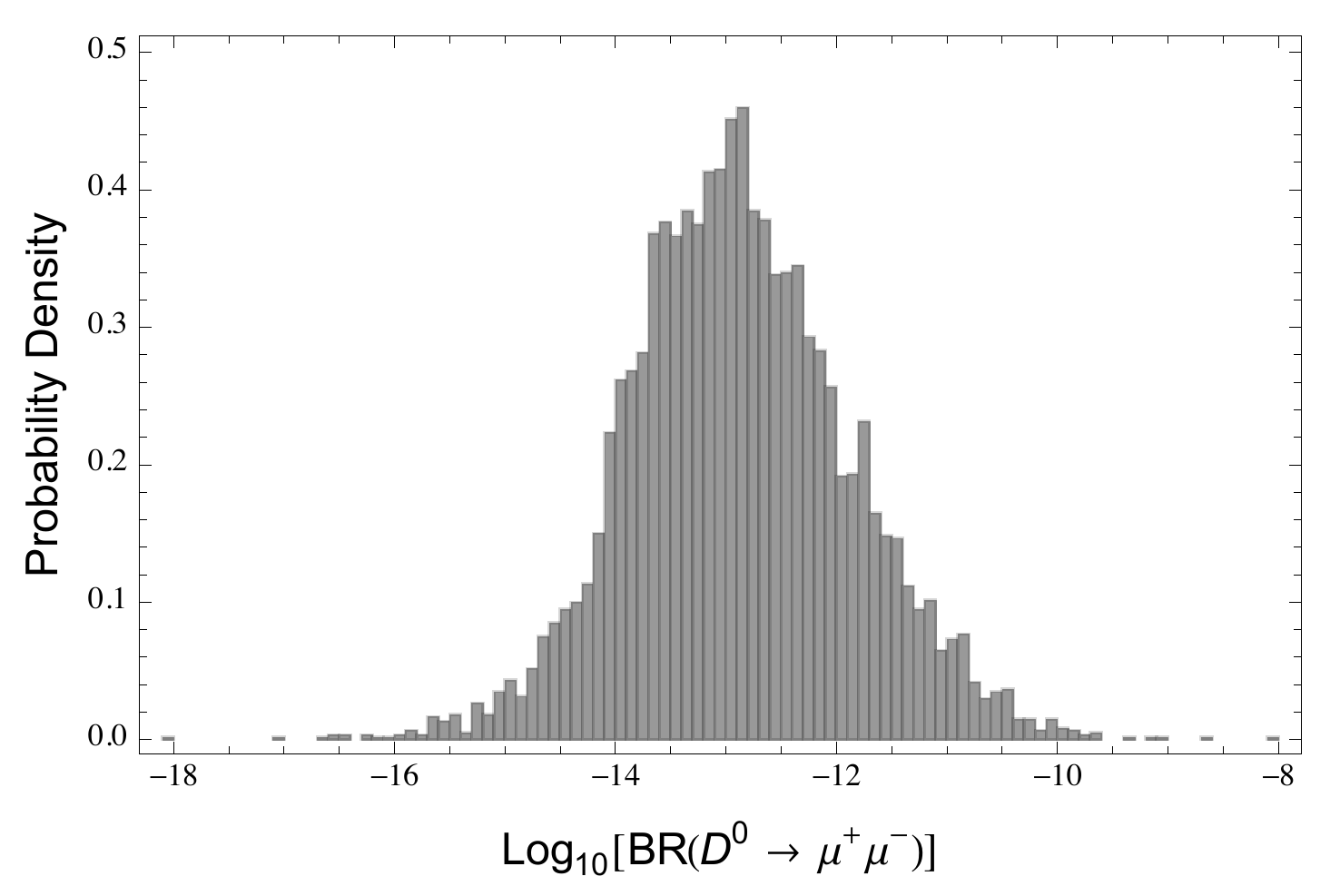}
\end{center}
\caption{Probability distribution of the branching fraction of $D^0\to \mu^+\mu^-$. Each bin is proportional to the density of parameter points that contribute to the range of values of ${\rm BR}(D^0\to \mu^+\mu^-)$ in that bin. For reference ${\rm Log_{10}[BR_{SM}^{LD}}(D^0\to \mu^+\mu^-)]=-12.09$.}
\label{fig:Dmumu}
\end{figure}

Contributions to $D^0\to \mu^+\mu^-$ are driven by tree level $V-A$ currents to the four fermion vertex $(\bar u c)_{V-A}(\mu^+\mu^-)_{V-A}$, which come primarily from the the mixing of the $Z^1$ and the $Z'$ with the $Z$ yielding the physical $Z_{1}$ state. The contributions of the other neutral gauge bosons are negligible to up to a few percent. The new contributions to the $V-A$ current can be parametrized through the following functions:
\begin{equation}
\Delta Y^{D}_{V-A}=-\sum_{m}\frac{g^{-}_{l}(m)g^{-}_{q}(m)}{4g^{2}_{SM}M_{m}},\;\;\;
\end{equation}
where 
$$g_{SM}=\left(\frac{G_{F}\alpha}{\sqrt{8}\pi\sin^{2}\theta_{W}}\right)^{1/2},$$
 and the functions $g^{\pm}_{l,q}$ parametrize the difference or sum between the left- and right-handed flavor non-diagonal couplings to the neutral gauge bosons, $Z\bar{q}_{iL,R}q_{jL,R}$ and $Zl^{+}_{L,R}l^{-}_{L,R}$, for $q_{i}=u,c$ and $l=\mu$. For quarks, these are obtained from Equations~(\ref{eq:UqZ}) and~(\ref{eq:UqA}) and are given by
\begin{equation}
g^{\pm}_{D}(m)=U^{q}_{L,eff}(m)_{2,1}\pm U^{q}_{R,eff}(m)_{2,1},
\end{equation}
where $m$ runs over the three neutral gauge boson mass eigenstates and the first KK mode of the photon. 
For leptons, using Equations~(\ref{eq:UlZ}) and~(\ref{eq:UlA}) we obtain
\begin{equation}
g^{\pm}_{\mu}(m)=U^{l}_{L,eff}(m)_{2,2}\pm U^{l}_{R,eff}(m)_{2,2}.
\end{equation}
This modifies the branching fraction to
\begin{eqnarray}
{\rm BR}(D^0\rightarrow \mu^+\mu^-)&=&\frac{1}{\Gamma_{D^0}}\frac{G_F^2}{\pi}\left(\frac{\alpha}{4\pi\sin^2(\theta_W)}\right)^2 f_D^2 m_\mu^2 m_{D^0} \sqrt{1-4\frac{m_\mu^2}{m_{D^0}^2}}|Y^D_{V-A}+\Delta Y^{D}_{V-A}|^2,\nonumber\\
Y^D_{V-A}&=&\sum_{j=d,s,b}V_{cj}^\ast V_{uj} \left(Y_0\left(x_j\right)+\frac{\alpha_s}{4\pi}Y_1\left(x_j\right)\right),
\end{eqnarray}
where $x_\mu=\mu^2/m_W^2$. The definitions of $Y_0\left(x_j\right)$ and $Y_1\left(x_j\right)$ can be found in \cite{Paul:2010pq}.

Figure~\ref{fig:Dmumu} clearly shows that enhancements of $O(10^1)-O(10^2)$ is possible over SM LD rates which is denoted by the darker region of the histogram. The LHCb now reports an upper bound of \cite{Aaij:2013cza}:
\begin{equation}
{\rm BR_{exp}}(D^0\rightarrow \mu^+\mu^-)< 6.2(7.6)\times 10^{-9}\,\,\, \textrm{ at 90\% (95\%) C.L.}.
\end{equation}
with 0.9fb$^{-1}$ of data. They can be expected to enhance this measurement by two or three orders of magnitude in the future.

Hence, this is a good channel to look for ND from the warped extra dimension even if no ND effects are seen in the analogous mode $B_s\to\mu^+\mu^-$ as we shall show later in Section \ref{sec:CorrBD}.

%%%%%%%%%%%%
\boldmath
\subsection{$D\to X_u \nu\bar{\nu}$}
\unboldmath
%%%%%%%%%%%%

\begin{figure}[h!]
\begin{center}
\includegraphics[width=8.5cm]{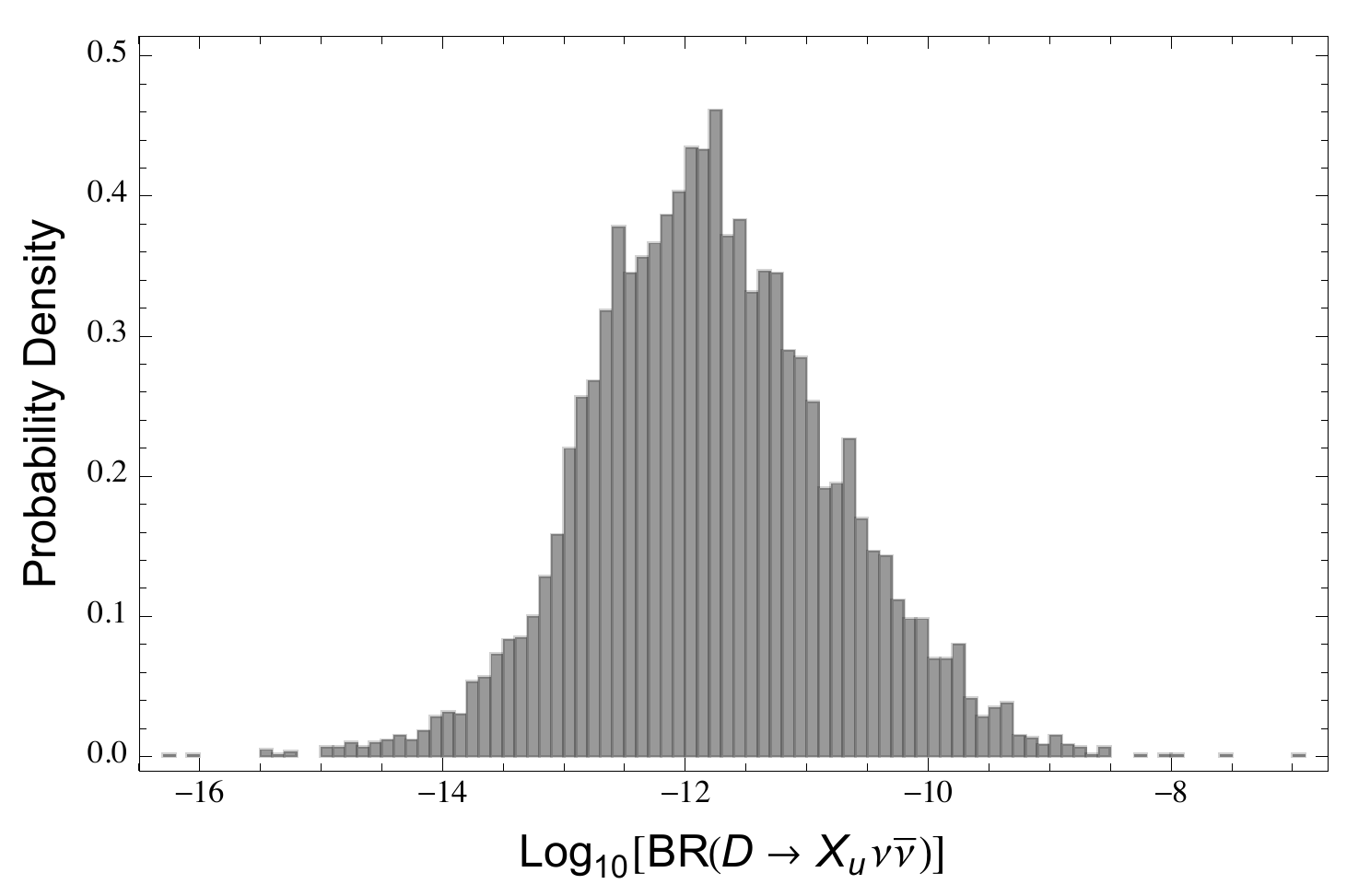}
\end{center}
\caption{Probability distribution of the branching fraction of $D\to X_u \nu\bar\nu$.}
\label{fig:Dnunu}
\end{figure}

The SM rate for $D\to X_u\nu\bar\nu$ is driven by SD operators and a detailed analysis with light quark loops is not expected to bring about much enhancements \cite{Buchalla:1993wq,Burdman:2001tf}. The new contributions to the $V-A$ and $V$ currents of this channel can be parametrized through the following functions:
\begin{eqnarray}
&&\Delta X^{D}_{V-A}=-\sum_{m}\frac{[g^{-}_{l}(m)+g^{+}_{l}(m)]g^{-}_{q}(m)}{8g^{2}_{SM}M_{m}},\nonumber\\
&&\Delta X^{D}_{V}=-\sum_{m}\frac{[g^{-}_{l}(m)+g^{+}_{l}(m)][g^{+}_{q}(m)-g^{-}_{q}(m)]}{8g^{2}_{SM}M_{m}}.\nonumber\\
\end{eqnarray}
The partial width is given by
\begin{eqnarray}
{\rm BR}(D\to X_u\nu\bar{\nu})=\frac{G^{2}_{F}m^{5}_{D}}{192\pi^{3}\Gamma_{D}}\left(\frac{\alpha}{4\pi\sin^{2}\theta_{W}}\right)^{2}\left(\left|X^{D}_{V-A}+\Delta  X^{D}_{V-A}+\frac{\Delta X^{D}_{V}}{2}\right|^{2}+\left|\frac{\Delta X^{D}_{V}}{2}\right|^{2}\right),
\end{eqnarray}
where $X^{D}_{V-A}$ come from SM \cite{Burdman:2001tf}. The vector contribution comes from ND only.

 It can be seen from Figure~\ref{fig:Dnunu} that ND from the warped extra dimension can bring about enhancements of $O(10^{5})-O(10^{6})$ over the SM rates in a large part of the parameter space. As for the case of the branching fraction of $D^0\to \mu^+\mu^-$, this is possible even when ND of such kind can only make negligible or no contributions to $B_s\to\mu^+\mu^-$. The source of such enhancement is the mixing of the ND gauge states with the SM $Z$ boson producing tree level FCNC. Numerically, almost all the enhancement is due to ND and comes from the $V-A$ FCNC.

%%%%%%%%%%%%
\boldmath
\subsection{$D\to X_ul^+l^-$}
\unboldmath
%%%%%%%%%%%%

\begin{figure*}
\begin{center}
\includegraphics[width=17cm]{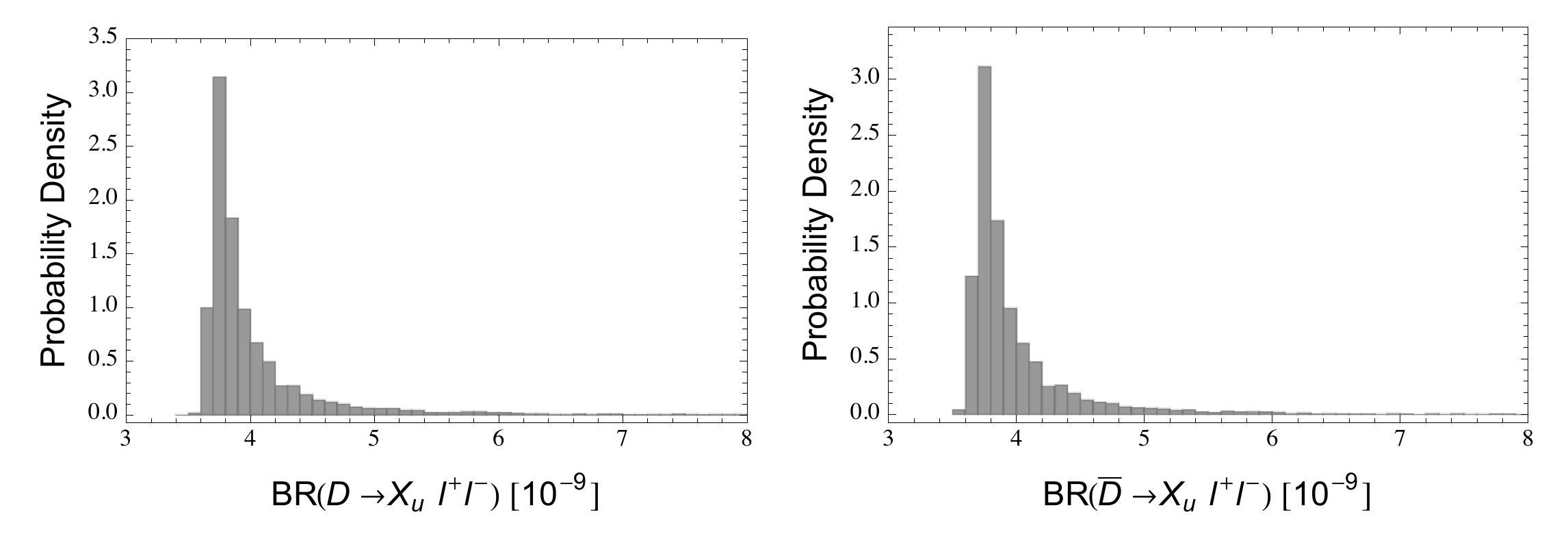}
\end{center}
\caption{Probability distribution of the branching fraction of $D\to X_u l^+l^-$.}
\label{fig:DXll}
\end{figure*}

As can be seen in Figure~\ref{fig:DXll}, the effects of ND in the branching fraction are quite negligible, $\sim 10\%$  in most of the parameter space. This is evident from the peak of the distribution lying on the SM value for the branching fraction. The additional contributions to $C_{9}$ and $C_{10}$ from tree-level neutral currents arising from the warped extra dimension are parametrized by:
\begin{eqnarray}
\Delta C_{9}&=&\left[\frac{\Delta Y^{D}_{V}+\Delta Y^{D}_{V-A}}{\sin^{2}\theta_{W}}-4\left(\Delta Z^{D}_{V}+\Delta Z^{D}_{V-A}\right)\right],\nonumber\\
\Delta C_{10}&=&-\left(\Delta Y^{D}_{V}-\Delta Y^{D}_{V-A}\right),
\end{eqnarray}
where
\begin{eqnarray}
&&\Delta Z^{D}_{V-A}=-\sum_{m}\frac{[g^{+}_{l}(m)-g^{-}_{l}(m)]g^{-}_{q}(m)}{16g^{2}_{SM}\sin^{2}\theta_{W}M_{m}} , \nonumber\\
&&\Delta Z^{D}_{V}=-\sum_{m}\frac{[g^{+}_{l}(m)-g^{-}_{l}(m)][g^{+}_{q}(m)-g^{-}_{q}(m)]}{8g^{2}_{SM}M_{m}}.
\end{eqnarray}
While $C_{10}$ is enhanced by orders of magnitude and $C_9$ gets significant contributions too, these tree level contributions fail to compete with the photon penguin contribution in $C_9$ coming from the SM. In addition to this, the branching fractions in these channels are dominated by long distance effects by orders of magnitude. Hence, new dynamics have no chance of showing up in the branching fractions of these channels. The expression for the differential branching fraction  in terms of the Wilson coefficients and the relevant references to existing literature can be found in Appendix \ref{app:BR}. 

A detailed study of $D\to P l^+l^-$, $P$ being a pseudoscalar, was done recently by the authors of \cite{Fajfer:2012nr}, where they extensively study the SM, and possible ND, contributions to both the branching fractions and asymmetries in these channels. The effects of models with a warped extra dimension in this channel was studied in \cite{Delaunay:2012cz}. We disagree with some of their statements. The largest effect that this kind of ND will have on this channel is not through its contribution to  the dipole operators in $c\to u \gamma$. In fact the dipole contributions to $C_7$ and $C_8$ hardly play a role in determining the size of the branching fraction of $D\to X_u l^+l^-$ even in the SM. It is the photon penguin contribution in $C_9$ which sets the stage for the SM SD contribution. The latter is not enhanced significantly by these models  over SM values. The real enhancement is seen in $C_{10}$ through tree level neutral currents which, however, can at most become comparable to the SM SD photon penguin contribution. 

It should be noted that it is also important to take into account final states with more than one hadron in addition to the lepton pair. Such decay modes are now being probed by LHCb. Their recent results put the upper limits on some exclusive modes with one \cite{Aaij:2013sua} or two \cite{Aaij:2013uoa} pions in the final state {\it below} the theoretical estimates of the SM long distance contribution to these channels \cite{Burdman:2001tf,Cappiello:2012vg}

%%%%%%%%%%%%
\section{Asymmetries in \boldmath $D\to X_{u} l^+l^-$}
\label{sec:Asym}
%%%%%%%%%%%%
\begin{figure*}
\begin{center}
\includegraphics[width=17cm]{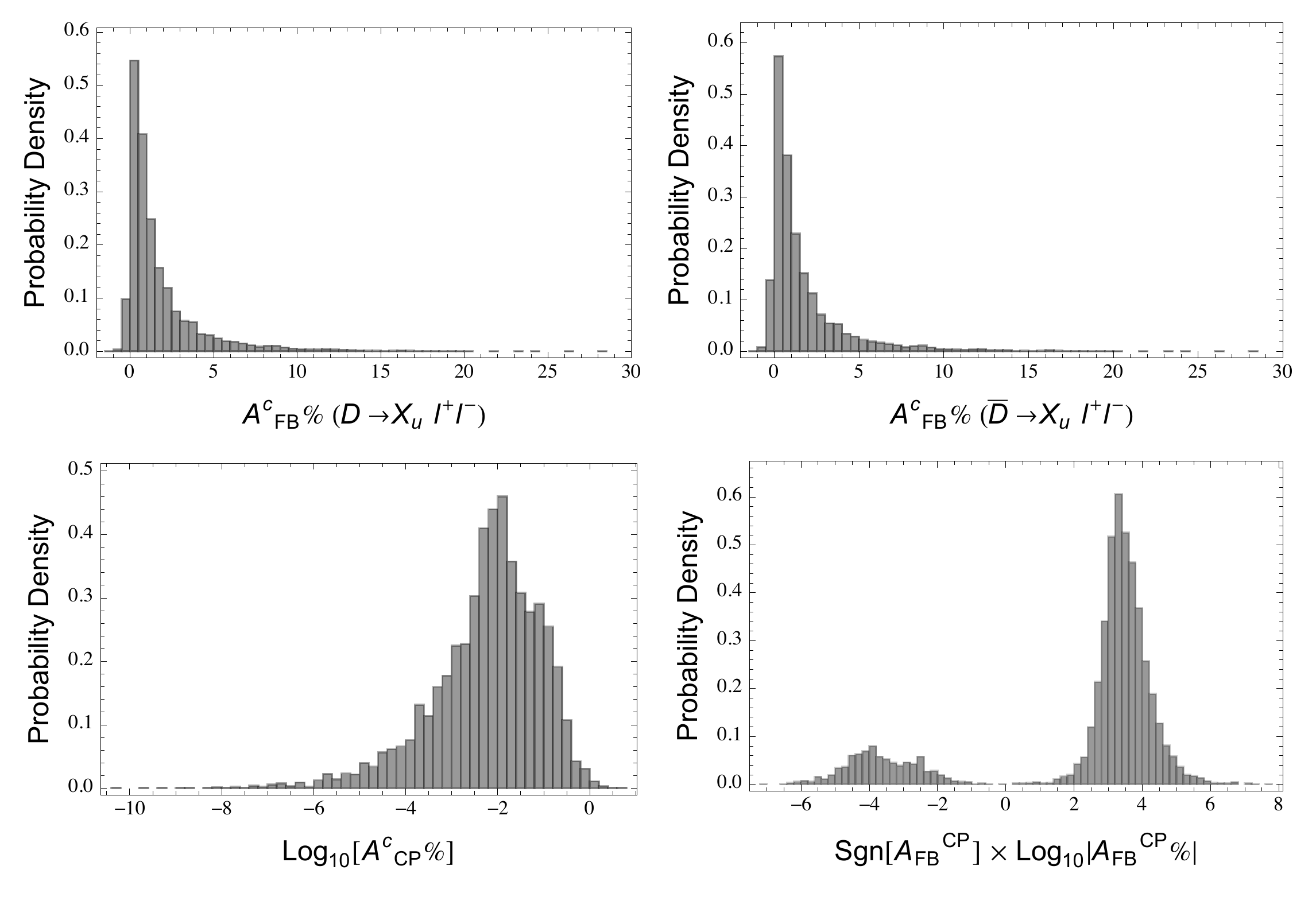}
\end{center}
\caption{Probability Distributions of the Asymmetries in $D\to X_u l^+l^-$.}
\label{fig:DXllAsym}
\end{figure*}

It is a different story altogether for the impact of ND on asymmetries in the $D\to X_{u} l^+l^-$ channel. The SM signatures in the asymmetries are extremely tiny. From Figure \ref{fig:DXllAsym}, it can be seen that ND intervention in the asymmetries is sizable or even large -- i.e.,  orders of magnitude more than what the SM can produce. The asymmetries are defined in Appendix \ref{app:ASYM}.

The forward backward asymmetry $A^c_{\rm FB}$ can be enhanced to even $O(5\%)$ in some parts of the parameter space. This enhancement can be understood from the enhancement of $C_{10}$ as $A^c_{\rm FB}$ depends directly on the magnitude of $C_{10}$. 

It should be noted that in the SM, and in the absence of CP violation (a good assumption within the SM for these channels), $A^c_{\rm FB}$ for the conjugate channels should have opposite signs. However, in Figure \ref{fig:DXllAsym}, it can be seen that the asymmetry for both the conjugate channels are of the same sign. This is a clear indication of the existence of CP asymmetry in these channels. 

The latter is made clear by the up to $O(1\%)$ CP asymmetry, $A^c_{\rm CP}$, that can be seen from Figure \ref{fig:DXllAsym}. Although CP asymmetry remains small for quite an insignificant fraction of the parameter space, there are possibilities of measurable CP asymmetry in a significant fraction of the parameter space too. This is not surprising as there are new phases coming from the new mass mixing matrices of the fermions affecting the neutral currents driven by new gauge states. 

As expected from large $A^c_{\rm FB}$ and sizeable $A^c_{\rm CP}$, the $A^{\rm CP}_{\rm FB}$ is enhanced by orders of magnitude in this scenario of ND. This asymmetry can be as high as $O(10^4\%)$ and can be both positive and negative although it does show a tendency of being positive in most of the parameter space. This can be attributed to the prevalence of positive $A^c_{\rm FB}$ over negative ones in a large part of the parameter space, but can also arise from $A^c_{\rm FB}$ being larger in $D$ than in $\bar D$\footnote{This can be concluded only since a very tiny part of the parameter space has $A^{\rm CP}_{\rm FB}$ less than 1\%}. 

 \begin{figure}
 \begin{center}
\includegraphics[width=18cm]{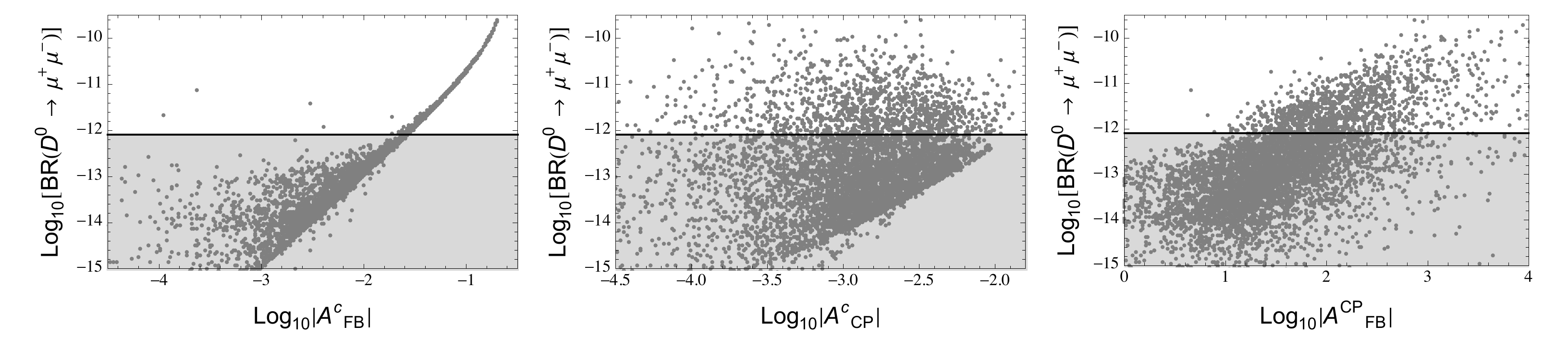}
 \end{center}
\caption{Correlations between asymmetries in $D\to X_u l^+l^-$ and ${\rm BR}(D^0\to\mu^+\mu^-)$. The region below the black line shaded in grey represents the part of the observable space where ND contribution to ${\rm BR}(D^0\to\mu^+\mu^-)$ is subdominant to SM LD contribution to the same.}
\label{fig:DobsEnh}
\end{figure}

%%%%%%%%%%%%%%%%%%%
\subsection{CPT invariance}
\label{CPTSYM}
%%%%%%%%%%%%%%%%

We have mostly focused on neutral charm mesons decays. 
Of course CPT invariance gives equality of masses and lifetimes of particles and antiparticles and affect meson oscillations.  
Yet this symmetry gives us much more; it gives equalities of {\em sub}-classes of $\Delta C =1$ 
decays defined by rescattering with mostly strong dynamics.  Often it is seen as only an academic tool from the world of quarks (\& gluons), 
since the measured final states consist of large numbers of hadrons. However, it is usable in 
in the decays of charm hadrons, since they produce only small numbers of hadrons in the final state -- in particular, for semi-leptonic decays and even 
more so for $D \to X_u l^+l^-$ vs. $\bar D \to \bar X_u l^- l^+$. Furthermore, one expects 
$\Gamma (D \to l^+l^- [\pi \pi/K\bar K]) \simeq \Gamma (\bar D \to l^+l^- [\pi\pi/K\bar K])$ especially in the resonance region where $(\rho/\omega/\phi \to l^+l^-)X_u$. 
There are subtle comments: 
\begin{itemize}
\item
CPT invariance tells us, in general : $\Gamma (D \to Y_u)= \Gamma (\bar D \to \bar Y_u)$; $Y_u/\bar Y_u$ include pairs of  $K\bar K$. 
\item 
We discuss classes of  WED models and their existence in inclusive $D$ decays with $l^+l^-$ in the final state due to short distance dynamics. Of course, one needs huge data to probe rare $D$ decays. 
\item 
On the other hand  the SM hardly produces a `background' in the {\em asymmetries} in $D \to X_u l^+l^-$. However, one needs more help: CPT invariance might be a usable tool.  It tells us that general equalities can be produced by `sizable' asymmetries in different regions of the dilepton invariant mass distribution which can shed light not only on the existence but maybe even the features of ND.

\end{itemize}
So far LHCb has produced a much lower limit on $D^0 \to \mu^+\mu^- \pi^+\pi^-$ \cite{Aaij:2013uoa} and $D^+_{(s)} \to \pi^+\mu^+\mu^-$ \cite{Aaij:2013sua} rates than predicted \cite{Burdman:2001tf,Cappiello:2012vg}. However, it is possible that LHCb might find a real signal for $D^0 \to \mu^+\mu^- K^+K^-$ -- and much later Belle II might measure $D^0 \to l^+l^- K^0 \bar K^0$ and even $D^0 \to l^+l^- \pi^0\pi^0$ to satisfy CPT invariance. Finally, one could find real sizable CP asymmetries and {\em correlations} between them.

%%%%%%%%%%%%
\section{Correlations between different \boldmath $\Delta C =1$ decay modes}
\label{sec:CorrDD}
%%%%%%%%%%%%

\begin{figure}
\begin{center}
\includegraphics[width=8.5cm]{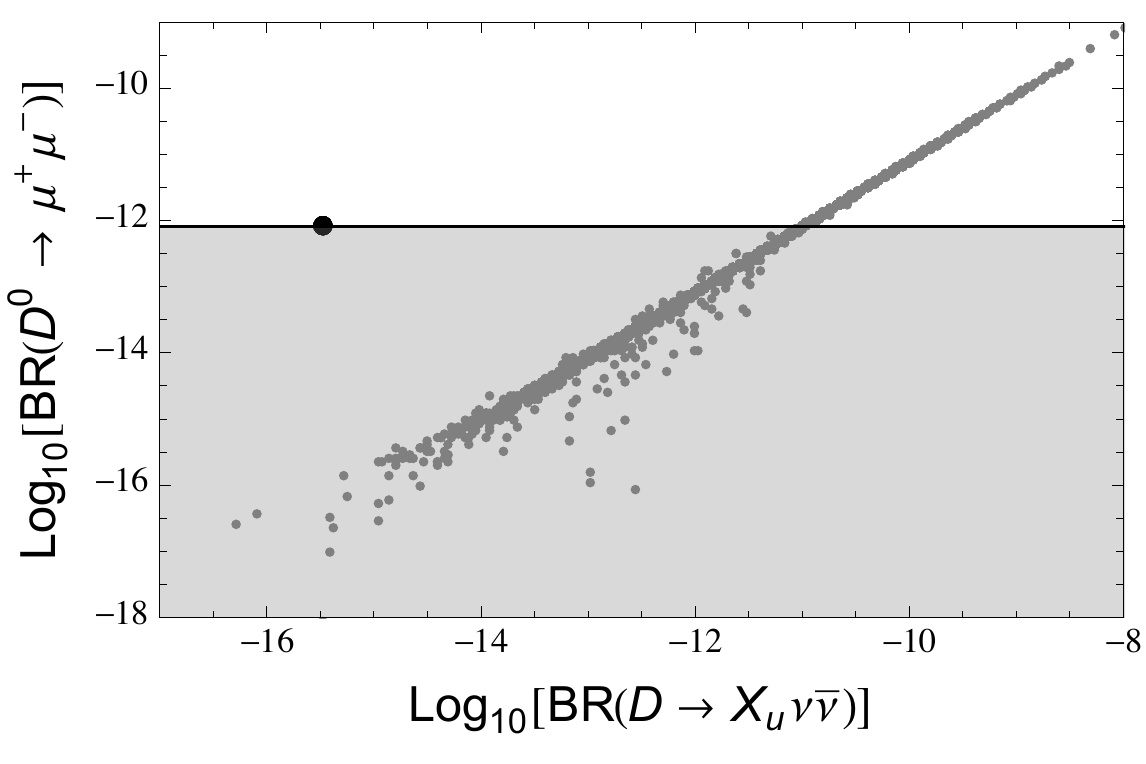}
\end{center}
\caption{Correlations between ${\rm BR}(D\to X_u\nu\bar\nu)$ and ${\rm BR}(D^0\to\mu^+\mu^-)$ . The black dot  represents the SM estimate of ${\rm BR}(D\to X_u\nu\bar\nu)$. The grey region represents the part of the parameter space where only ${\rm BR}(D\to X_u\nu\bar\nu)$ gets enhanced above SM estimated while ND in ${\rm BR}(D^0\to\mu^+\mu^-)$ remains subdominant to SM.}
\label{fig:Dmumununu}
\end{figure}

It is instructive to take a look at the correlation between the observables in different decay modes. In Figure \ref{fig:DobsEnh} we look at the correlation between the asymmetries in $D\to X_u l^+l^-$ and $D^0\to\mu^+\mu^-$. The grey band denotes the part of the parameter set that fails to overcome the SM LD contribution to $D^0\to\mu^+\mu^-$. Therefore the points in the white region of each of the graphs represent enhancements to {\em both} the branching fraction of $D^0\to\mu^+\mu^-$ and the asymmetries in $D\to X_u l^+l^-$ that, if observed, would hint at the presence of ND. All the asymmetries show possibilities of enhancements to large values while the branching ratio of $D^0\to\mu^+\mu^-$ remains large and distinguishable from SM LD contributions. The only two pairs of observables that are clearly correlated are the branching fraction of $D^0\to\mu^+\mu^-$ and $A^c_{\rm FB}$ since both depend purely on the size of $C_{10}$.

The case for the the correlations between the branching fractions of $D^0\to\mu^+\mu^-$ and $D\to X_u\nu\bar\nu$ is shown in the plot on the right in Figure \ref{fig:Dmumununu}. In the grey region the branching fraction of $D^0\to\mu^+\mu^-$ lies shrouded in SM long distance dynamics. We see here that there are significant parts of the parameter space in which simultaneous enhancements to the branching fraction of both $D^0\to\mu^+\mu^-$ and $D\to X_u\nu\bar\nu$ are possible due to the tight correlation between the two observables. It can also be seen that even if ND shows up in $D\to X_u\nu\bar\nu$, it can fail to overcome the SM contribution to $D^0\to\mu^+\mu^-$ in a large part of the parameter space. However, it should be kept in mind that in the latter regime, it will be quite difficult for experiments to measure the branching fraction of $D\to X_u\nu\bar\nu$ due to its small size.

%%%%%%%%%%%%
\section{Correlations between strange, charm and beauty}
\label{sec:CorrBD}
%%%%%%%%%%%%
\begin{figure}
\begin{center}
\includegraphics[width=8.5cm]{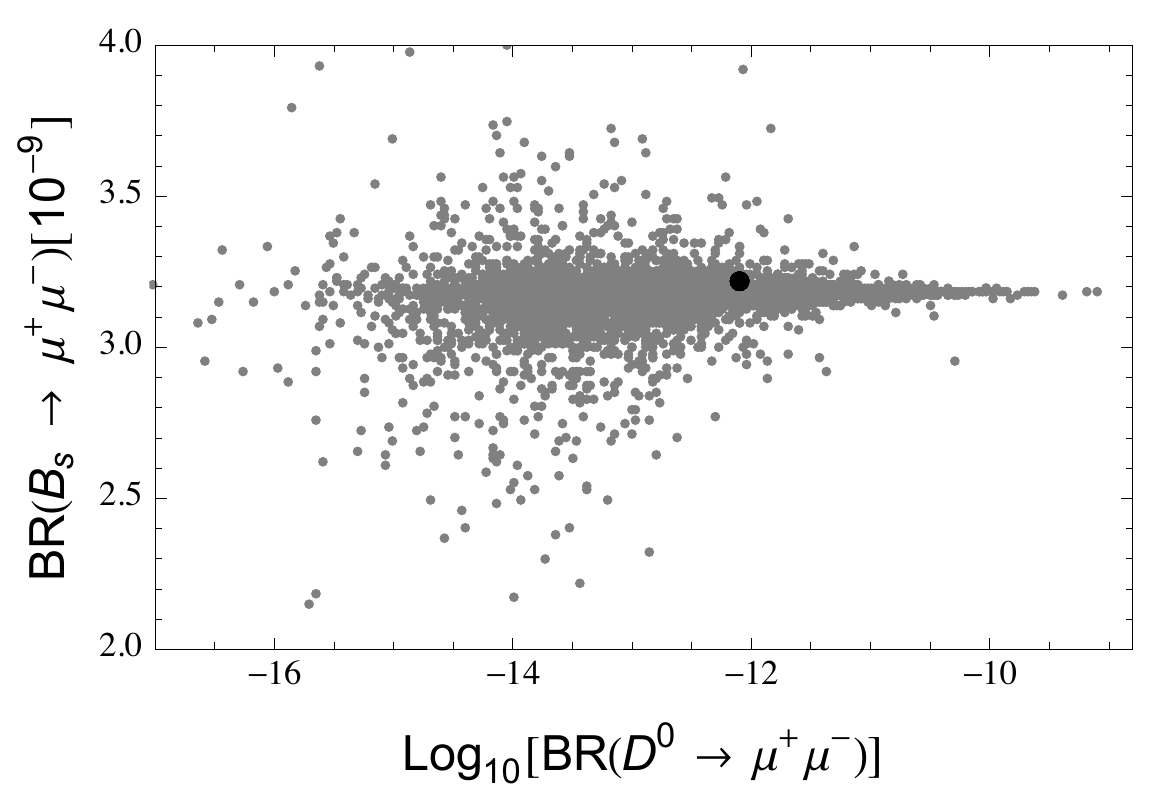}
\end{center}
\caption{Correlations between ${\rm BR}(B_s\to\mu^+\mu^-)$ and ${\rm BR}(D^0\to\mu^+\mu^-)$ . The black dot  represents the SM LD estimate of ${\rm BR}(D^0\to\mu^+\mu^-)$. The y axis is approximately within the 1$\sigma$ values for the current LHCb \cite{Aaij:2013aka} and CMS \cite{Chatrchyan:2013bka} results for ${\rm BR}(B_s\to\mu^+\mu^-)$.}
\label{fig:BDmumu}
\end{figure}

\begin{figure*}
\begin{center}
\includegraphics[width=17.5cm]{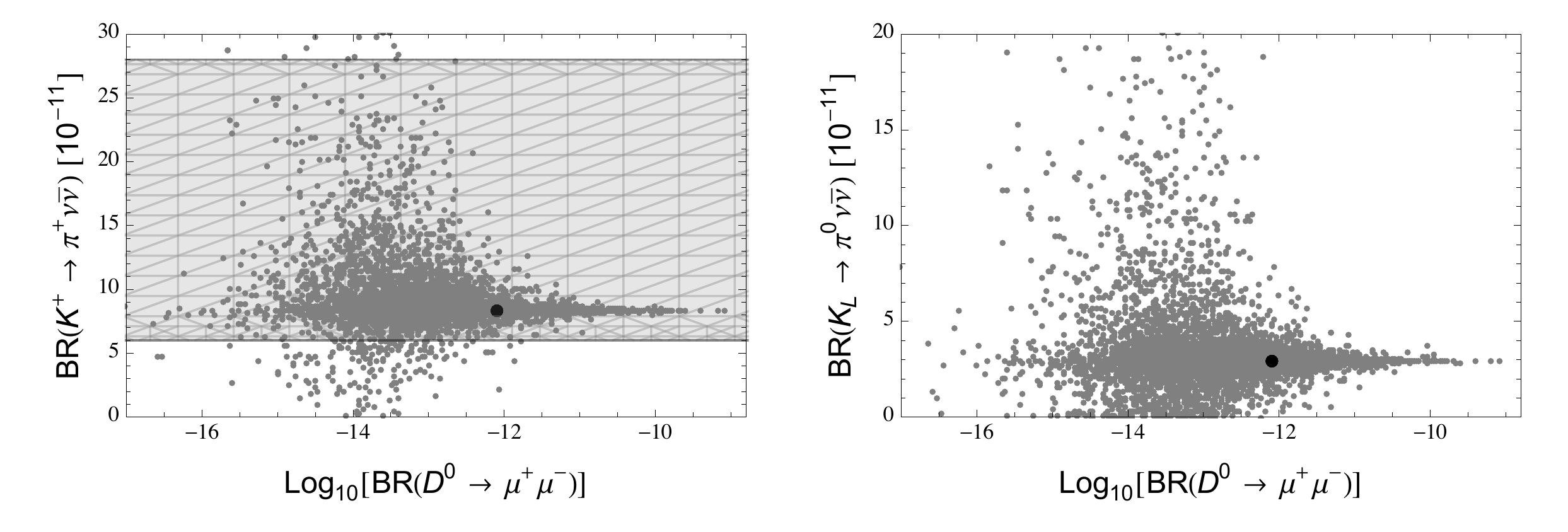}
\end{center}
\caption{Correlations between ${\rm BR}(K^+\to\pi^+\nu\bar\nu)$, ${\rm BR}(K_L\to\pi^0\nu\bar\nu)$ and ${\rm BR}(D^0\to\mu^+\mu^-)$ . The black dot  represents the SM LD estimate of ${\rm BR}(D^0\to\mu^+\mu^-)$ along with the SM estimates for ${\rm BR}(K^+\to\pi^+\nu\bar\nu)$ and ${\rm BR}(K_L\to\pi^0\nu\bar\nu)$. The grey band on the left plot is the $1\sigma$ experimental bounds on ${\rm BR}(K^+\to\pi^+\nu\bar\nu)$.}
\label{fig:KpinunuCorr}
\end{figure*}

There are possibilities of nonuniversal scenarios of flavour dynamics that can manifest themselves and leave different signatures in different flavour sectors. Manifestations of ND also need to dominate over SM contributions to flavour observables to be seen in flavour dynamics. While this is not possible in some flavour observable, some others, specially from charm dynamics, are ripe for such occurrences. A correlation study between $\Delta S=1$ and $\Delta B=1$ observables was done in \cite{Blanke:2008yr}. It is quite instructive to also compare flavour dynamics in two different sectors.

In Figure \ref{fig:BDmumu} we show the correlation between ${\rm BR}(B_s\to\mu^+\mu^-)$ and ${\rm BR}(D^0\to\mu^+\mu^-)$. The values on the y axis lie approximately within the 1$\sigma$ values from the current experimental measurement of the branching fraction of $B_s\to\mu^+\mu^-$ announced recently by LHCb \cite{Aaij:2013aka} and CMS \cite{Chatrchyan:2013bka}:
\begin{equation}
{\rm BR}(B_s\to\mu^+\mu^-)=\left\{
\begin{array}{ll}
(2.9^{+1.1}_{-1.0}\;{\rm (stat)}\;^{+0.3}_{-0.1}\;{\rm(syst)})\times 10^{-9}&\;\;{\rm LHCb}\vspace{0.1cm}\\
(3.0^{+1}_{-0.9})\times 10^{-9}&\;\;{\rm CMS}\\
\end{array}
\right.
\end{equation}
to be compared to the SM estimate\cite{Buras:2012ru}
\begin{equation}
{\rm BR_{SM}}(B_s\to\mu^+\mu^-)=(3.54\pm0.30)\times10^{-9}\; .
\end{equation}

There is not much correlation between the two observables. However, it is clear that even if the experimental errors are reduced and the branching fraction of $B_s\to\mu^+\mu^-$ is very close to the SM expectation and ND effects are indiscernible in that channel, large enhancements can show up in the branching fraction of $D^0\to\mu^+\mu^-$ and even dominate over the SM LD contributions to the same. This stems from the fact that SM contributions to $B_s\to\mu^+\mu^-$ (like in most of beauty dynamics) are quite large. Hence, ND can hardly make orders of magnitude enhancement there. The enhancements to $B_s\to\mu^+\mu^-$ can clearly be seen here to be of $O(10\%)$ while in the same parameter space it is orders of magnitude for charm. Hence, new dynamics can manifest itself in charm while keeping a low profile in beauty and vice versa, although the latter is a more difficult situation to deal with experimentally. We advocate the study of ND in charm lest nature chooses the former.

In \cite{Blanke:2008yr,Blanke:2008zb}, it was shown that the observables in the strange and beauty sectors are anti-correlated and can show enhancements in complementary parts of the parameter space. However, even in the $\Delta S=1$ observables studied in the kaon sector, namely the branching fractions of $K_L\to\mu^+\mu^-$, $K^{(+)}_L\to\pi^{0(+)}\nu\bar\nu$ and $K_L\to\pi^0l^+l^-$ the enhancements are modest and of at most an order of magnitude. The last two are compared in Figure \ref{fig:KpinunuCorr} with ${\rm BR}(D^0\to\mu^+\mu^-)$. The experimental limits for these two kaon decay modes are \cite{Artamonov:2008qb,Ahn:2009gb}:
\begin{eqnarray}
{\rm BR}(K^+\to\pi^+\nu\bar\nu)&=&1.7\pm 1.1\times 10^{-10},\\
{\rm BR}(K_L\to\pi^0\nu\bar\nu)&<&2.6\times 10^{-8}\;\;{\textrm{at 90\% CL}},
\end{eqnarray}
to be compared to the SM values of \cite{Mescia:2007kn}

\begin{eqnarray}
{\rm BR}(K^+\to\pi^+\nu\bar\nu)&=&7.83\pm 0.82\times 10^{-11},\\
{\rm BR}(K_L\to\pi^0\nu\bar\nu)&=&2.49\pm0.39\times 10^{-11}.
\end{eqnarray}

While it is true that if the experimental value stays close to the central value for ${\rm BR}(K^+\to\pi^+\nu\bar\nu)$, this kind of ND cannot make large enhancements to ${\rm BR}(D^0\to\mu^+\mu^-)$ it is also true in some parts of the parameter space, even if ND leaves very small signatures in the former, the latter can be enhanced by orders of magnitude in a large part of the parameter space.  With experimental limits being quite close to the SM, the case of ND suffers the same fate as in beauty: there is a possibility of enhancements for ND but it lies shrouded in the shadows of the SM. We emphasise again, the case of charm is very different. Where ND can only leave modest effects in strange and beauty, it can leave a severe impact in charm, sometimes orders of magnitude beyond the reach of the SM.

%%%%%%%%%%%%
\section{Analysis of the parameter space}
\label{sec:AnaParam}
%%%%%%%%%%%%
\begin{figure*}
\begin{center}
\includegraphics[width=17cm]{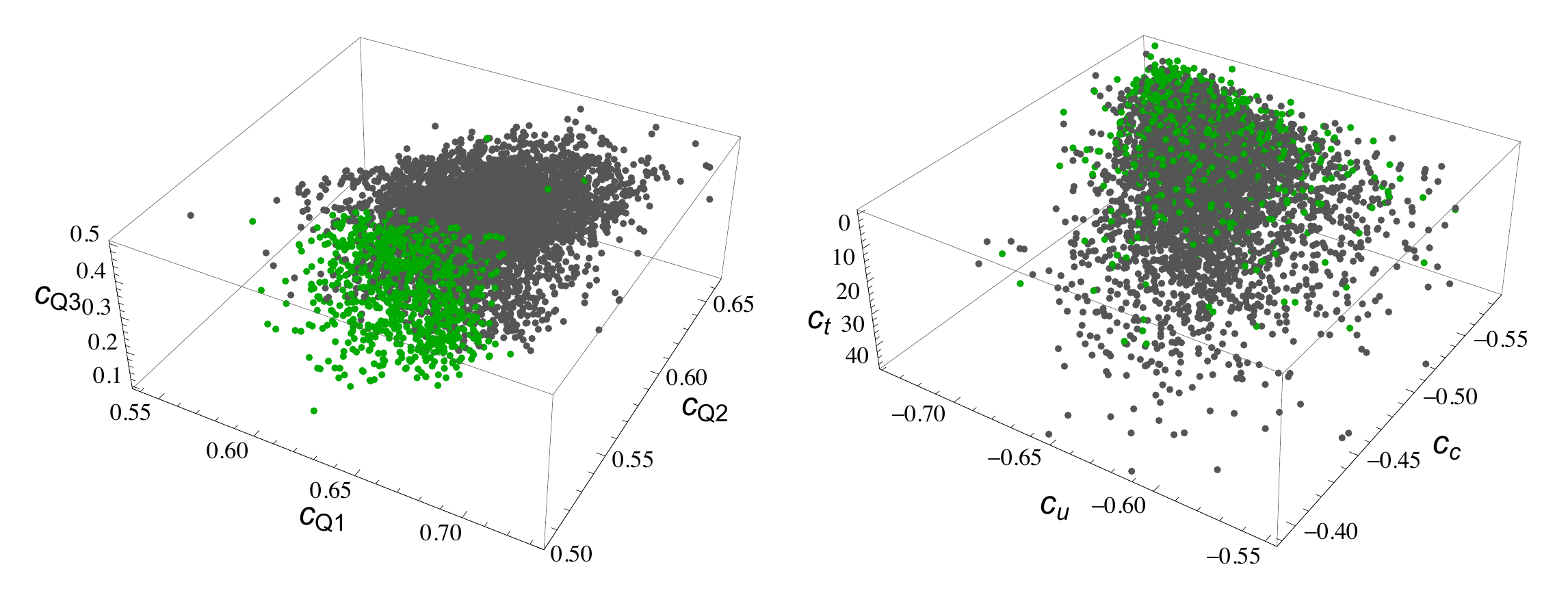}
\end{center}
\caption{The bulk mass parameters of the three families of quarks and their correlation to enhancements in ${\rm BR}(D^0\to\mu^+\mu^-)$. The figure on the left shows the bulk mass parameters for the left-handed quark doublets and the one on the right shows those for the right-handed singlet quarks. The green (lighter) points represent the part of the parameter space that can generate enhancements to ${\rm BR}(D^0\to\mu^+\mu^-)$ beyond the SM LD estimates.}
\label{fig:DBulk}
\end{figure*}
While we see enhancements in quite a few observables in rare charm decays, it might be instrumental to ask at this point whether such enhancements lead us to a preference for any particular part of the parameter space for the generation of observable new dynamics. In Section \ref{sec:flavWD} we displayed the distribution of the Yukawa couplings that was used in this analysis. In Figure \ref{fig:DBulk} we take another look at the space of the bulk mass parameters in the light of ND contributions to ${\rm BR}(D^0\to\mu^+\mu^-)$ beyond the SM LD estimates.

The green (lighter) points represent the part of the parameter space in which we see enhancements to ${\rm BR}(D^0\to\mu^+\mu^-)$ beyond the SM LD estimates. We clearly see, from the figure on the left, that there is a strong preference for the bulk mass parameter of the second family of left-handed SM quark doublets to be at lower values close to 0.5 when such enhancements occur, while there are no clear preferences for the values of the other two quark doublet bulk mass parameters. The figure on the right also shows that such enhancements prefer lower values of $c_t$, the bulk mass parameter of the right-handed top quark. No direct correlation can be drawn between the ND contributions and $c_t$ for even moderately large values of the latter. Such dependence on fermionic bulk mass parameters of the other right-handed quarks are absent as they are in general of $O(1)$. Although very large values of $c_t$ are considered unnatural, we have kept them in our analysis as they are allowed by the constraints we used to determine our parameter space and we did not wish to impose any additional arguments of naturalness on the parameters. 
 
The Yukawa couplings do not show any trends and the parameter points generating enhancements in ${\rm BR}(D^0\to\mu^+\mu^-)$ beyond the SM LD estimates are more or less randomly distributed over the entire parameter space. 

%%%%%%%%%%%%
\section{Comments on \boldmath $\Delta C = 2$ processes.}
\label{sec:DC2}
%%%%%%%%%%%%

To complete an analysis of the effects of any model on charm dynamics it is essential to address not only its effects in $\Delta C = 1$ processes, but also $\Delta C = 2$ dynamics, i.e., oscillations of the neutral charm meson system. Besides the importance of oscillations in providing the crucial ``other'' amplitude necessary for indirect CP violation, its parameters are measured well enough now \cite{Aubert:2007wf,Staric:2007dt,Abe:2007rd,Asner:2010qj,Aaij:2012nva} to raise the question of whether it can constrain parameter spaces of models of new dynamics. However, the size of the SM contribution to charm meson oscillations is debatable \cite{Bigi:1997xh,Bigi:2000wn,Bianco:2003vb,Bobrowski:2010xg,Lenz:2010pr,Falk:2001hx,Falk:2004wg}. In fact, the theoretical uncertainties surrounding the relative size of the SM vs. ND contributions to oscillations \cite{Blum:2009sk,Blanke:2007ee,Bigi:2009df} make it very difficult to conclude which one is the leading contribution or if both compete to deliver what we observe experimentally. Charm oscillations have been studied in models of warped extra dimensions too \cite{Bauer:2009cf} and can be shown to have large effects. 

There is another subtlety involved here. $D^0\to\mu^+\mu^-$, shares topologies with $D^0 - \bar{D}^0$ oscillations through the box diagram and possible neutral electroweak gauge boson exchange in the presence of new degrees of freedom of this nature. Hence, one can argue that a large enhancement to one channel would imply a large enhancement to the other \cite{Golowich:2009ii}. Even though this is in general a good argument, it does not always hold true \cite{Paul:2010pq}.

The case for the model in this study is quite different. Firstly, only the electroweak gauge bosons and their KK partners contribute to $D^0\to\mu^+\mu^-$. These gauge bosons while also being involved in $D^0 - \bar{D}^0$ oscillations through similar topologies, do not turn out to be the dominant contribution. $D^0 - \bar{D}^0$ oscillations also receive contributions from KK gluons and, like in the case of oscillations in the kaon sector, and unlike oscillations in the beauty sector, these are the contributions which dominate in the oscillation parameters. While we have done the calculations to check the same, we feel that the uncertainties surrounding the SM contribution to oscillations and the subtleties of their cancellation or reinforcement of the ND contributions demand that we deal with it in detail in a possible future work. Suffice it to say here that, in the light of theoretical uncertainties, there isn't a specific prescription for ruling out parameter space  we study here on the basis of the contributions of ND to charm oscillations.

%%%%%%%%%%%%
\section{Summary}
\label{sec:Sum}
%%%%%%%%%%%%
We claim that as long as models with a warped extra dimension have a flavour structure akin to what we have studied, it will contribute to charm changing neutral currents significantly. Here we have considered a non-ad hoc flavour structure that is not ``tuned'' to give effects in the {\em up}-type quark sector. Moreover, we have used a parameter space whose effects have already been studied in \cite{Blanke:2008yr,Blanke:2008zb} for effects in $B$ and $K$ physics. Hence it gives us a comprehensive picture of the effects of these kinds of flavour structures in the dynamics of {\em both} {\em up}- and {\em down}-type quarks.
 
In many models with a warped extra dimension it is possible to keep low KK scales using different theoretical technologies. Yet, one is led to wonder what will happen if we fail to discover these low KK scales, i.e., whether nature has ``heavier'' plans for us. The most simplistic way one can argue the KK scale dependence in the observables we have studied is to state that the amplitudes depend roughly on the inverse square of the KK mass scale and hence the branching fractions and the asymmetries depend on the same to the fourth power. The KK mass scale that we have used is about $2.45$ TeV. It would be quite justifiable to say that if the KK scale really lies at around 5 TeV or more\footnote{We are assuming the reality of the existence of a warped extra dimension.}, direct searches at LHC will start having a problem seeing new degrees of freedom. Yet the effects in charm dynamics would at best be lowered by an order of magnitude. Considering what the numbers tell us, charm dynamics would still be in the game for showing ND effects even if  direct searches at the LHC and possibly, both beauty and kaon dynamics would be out of the game within the ambits of such a ND scenario. Suffice it to say that charm has been both cursed and blessed by tiny SM signatures.

We also, in general, disagree with the conclusion in \cite{Delaunay:2012cz} that the ``only'' place the models with a warped extra dimension can show their effect are in CP asymmetries in $D\to X_u\gamma$ besides $\Delta A_{\rm CP}$\footnote{For detailed studies of the connection between these two observables cf. \cite{Isidori:2012yx,Lyon:2012fk}.}. We think we have convincingly argued in this paper that models of this kind can have large effects well beyond the SM estimates in multiple rare decays including $D^0\to\mu^+\mu^-$, $D\to X_u\nu\bar\nu$ and asymmetries in $D\to X_u l^+l^-$ which can be within experimental reach in the near future. Also, we propose a study of higher multiplicity hadronic decays of the charmed meson both experimentally and theoretically, although a lot of tools need to be developed for these.  

In this work we have shown that:
\begin{itemize}
\item These kind of flavour structure can leave large effects in charm, sometimes orders of magnitude larger than what the SM can generate.
\item While these models leave moderate effects in beauty and strange dynamics, the effects in charm need not be moderate. This can be achieved even without giving the {\em up}-type quark sector a special dynamical advantage. 
\item The effects in charm dynamics are not tightly correlated with those in beauty and strange, i.e., even when this kind of ND can leave negligible contributions to both beauty and strange dynamics, it can leave very large contributions to charm dynamics. 
\item These tree level effects coming from this class of models can be larger than the loop-level enhancement that we saw in the LHT-like models for the same observables.
\end{itemize}

While it continues to be true that accessing rare charm dynamics is statistically challenging and theoretically not well understood even at the level of the SM contributions, we have made some strong cases for ND intervention way beyond what the SM can generate. These effects not only lead to the observables coming within reach of current experiments and future super flavour factories, but also sidestep the problem of determining SM contributions theoretically due to their large size. 

Finally, we want to state that indirect evidences for ND are based on flavour dynamics beyond the SM 
in general. However we do not like to go to a shopping mall to find anything that is just beyond the SM 
flavour dynamics; we greatly prefer to think about flavour dynamics that originates from a motivation to find solutions to the hierarchy problem or challenge of 
the SM. Previously we have worked with LHT to deal with models of one class, now we have done the same for the warped extra dimension, {\it \`a la}, the class 
of Randall-Sundrum models. Now we shall wait and hope that nature conspires on our side.

%%%%%%%%%%%%
\section{Acknowledgement}
%%%%%%%%%%%%
We would like to express our gratitude to the authors of \cite{Blanke:2008zb,Blanke:2008yr} for sharing the parameter sets used in these works which helped us build correlations between the three flavour sectors. We would like to especially thank Monika Blanke and Stefania Gori for clarifying our queries about their work. We express our gratitude to Jorge de Blas Mateo for his `game changing' criticism. One of us, AP, would like to thank the Physics Department of the University of Notre Dame du Lac for providing computational resources and Rahul Sinha  for hosting him at the Indian Institute of Mathematical Science during the initial stages of this work. We acknowledge partial support from the European Research Council under the European Union's Seventh Framework 
Programme (FP/2007-2013) / ERC Grant Agreement n.~279972 and from the NSF under the Grant No. PHY-0807959/PHY-1215979. The work of ADLP was supported by the National Science and Engineering Research Council of Canada (NSERC).

\appendix
%%%%%%%%%%%%
\section{\boldmath $D^\pm\to X_u l^+l^-$}
\label{app:BR}
%%%%%%%%%%%%

The SM SD differential branching fraction for the inclusive process $D\to X_u l^+l^-$ is given by \cite{Paul:2011ar}:
\begin{eqnarray}
\nonumber &&\frac{d}{d\hat{s}}{\rm BR}_{\rm SM}^{\rm SD}\left(D\to X_u l^+l^-\right)=\\&&\frac{1}{\Gamma _D}\frac{G_F^2\alpha ^2m_c^5}{768\pi^5}(1-\hat{s})^2
\Bigg[\left(\left|C_9(\mu)\right|^2+\left|C_{10}(\mu)\right|^2\right)(1+2\hat{s})
+12 \text{ Re}(C_7(\mu)C^*_9(\mu))+4\left(1+\frac{2}{\hat{s}}\right)\left|C_7(\mu)\right|^2\Bigg],
\end{eqnarray}
with
\begin{eqnarray}
\nonumber \hat{s}=\frac{(p_{l^+}+p_{l^-})^2}{m_c^2}.
\end{eqnarray}

We set $\mu=m_c=1.2$ GeV. Integrating over $\hat{s}$ gives us the total decay rate. One has to be careful about not picking up the infrared divergence in the differential decay rate. We made an infrared cut on $\hat{s}$ at about an invariant dilepton momentum of $20$ MeV. The definitions of the Wilson operators and the form of the coefficients can be found in \cite{Paul:2011ar}. Models with a warped extra dimension can leave their impact in $C_9$ and $C_{10}$ trough tree level contributions to $Y(x)$ and $Z(x)$ defined in Section~\ref{sec:RareDecays}. 

%%%%%%%%%%%%
\section{Asymmetries in \boldmath $D^\pm\to X_u l^+l^-$}
\label{app:ASYM}
%%%%%%%%%%%%

Although the branching ratios are dominated by long distance physics, the asymmetries are sensitive to mostly short distance physics. Hence asymmetries are good observables for the discovery of ND as the latter is expected to bring enhancements to the short distance dynamics. There are three asymmetries that can be probed in this channel which come solely from $\Delta C=1$ currents \cite{Paul:2011ar}. 

The forward-backward asymmetry $A_{\rm FB}^c$ is tiny in the SM, $O(10^{-6})$, and the impact of new dynamics here can be quite large. The normalized forward-backward asymmetry is defined from the double differential decay rate as
\begin{eqnarray}
A^c_{\rm FB}(\hat{s})=\frac{\int_{-1}^{1}\left[\frac{d^2}{d\hat{s}dz}\Gamma(D^\pm\to X_ul^+l^-)
 \right]{ \rm sgn}(z)dz}
{\int_{-1}^{1}\left[\frac{d^2}{d\hat{s}dz}\Gamma(D^{\pm}\to X_ul^+l^-)\right]dz}\;.
\end{eqnarray}
After performing the integral over the angular distribution we obtain
\begin{eqnarray}
\nonumber &&A^c_{\rm FB}(\hat{s})=\frac{-3\left[ \Re (C_{10}^*(\mu) C_9(\mu))\hat{s} +2 \Re (C_{10}^*(\mu) C_7(\mu))\right]}{(1+2\hat{s})\left(\left|C_9(\mu)\right|^2+\left|C_{10}(\mu)\right|^2\right)+4\left|C_7(\mu)\right|^2\left(1+\frac{2}{\hat{s}}\right)+12\Re \left(C_7(\mu) C_9^*(\mu)\right)}\;.\\
\end{eqnarray}

Enhancements to both $C_9$ and $C_{10}$ can manifest themselves as sizable $A^c_{\rm FB}$. 

The CP asymmetry $A_{\rm CP}^c$ is of $O(10^{-4})$ in the SM and it is defined as
\begin{eqnarray}
A^c_{\rm CP}(\hat{s})=\frac{\frac{d}{d\hat{s}}\Gamma(D^+\to X_u l^+l^-)-\frac{d}{d\hat{s}}\Gamma(D^-\to X_{\bar{u}} l^+l^-)}{\frac{d}{d\hat{s}}\Gamma(D^+\to X_u l^+l^-)+\frac{d}{d\hat{s}}\Gamma(D^-\to X_{\bar{u}} l^+l^-)}\;.\nonumber\\
\end{eqnarray}
Integrating over the invariant dileptonic mass we get the total CP asymmetry $A^c_{\rm CP}$

The CP asymmetry in the forward-backward asymmetry $A_{\rm FB}^{\rm CP}$ can show very large contributions from new dynamics. The SM contribution to this asymmetry is of $O(10^{-5})$. Since this asymmetry is sensitive to phases in the Wilson coefficients, it is open to enhancements by ND. The normalized difference in the forward-backward asymmetry in $D\to X_u l^+l^-$ and $\bar{D}\to X_{\bar{u}} l^+l^-$ is defined as \cite{Buchalla:2000sk}
\begin{eqnarray}
A^{\rm CP}_{\rm FB}(\hat{s})=\frac{A^c_{\rm FB}(\hat{s})+A^{\bar{c}}_{\rm FB}(\hat{s})}{A^c_{\rm FB}(\hat{s})-A^{\bar{c}}_{\rm FB}(\hat{s})}\;.
\end{eqnarray}
In the limit of CP symmetry, $A^c_{\rm FB}(\hat{s})$ and $A^{\bar{c}}_{\rm FB}(\hat{s})$ have to be exactly equal in magnitude but with an opposite sign \cite{Kruger:1997jk,Choudhury:1997xa}. As the forward-backward asymmetry is defined in terms of the positive 
anti-lepton, $A^c_{\rm FB}(\hat{s})$ and $A^{\bar{c}}_{\rm FB}(\hat{s})$ have opposite signs. $A^{\rm CP}_{\rm FB}(\hat{s})$ is sensitive to the phase in $C_7$,  $C_9$ and $C_{10}$. The SM offers phases only in $C_7$ and $C_9$ in 
$D\to X_u l^+l^-$ and none in $C_{10}$. Hence the integrated asymmetry turns out to be very small.
\begin{equation}
\int A^{\rm CP}_{\rm FB}(\hat{s}) d\hat{s}=A^{\rm CP}_{\rm FB}\sim3\times 10^{-5}.
\end{equation}

To make the study of these asymmetries ``clean'' both theoretically and experimentally, cuts in the dilepton mass distribution can be made around the $\rho, \omega$ and $\phi$ resonances. Our previous study showed that making such cuts left our results unaltered proving that long distance dynamics, especially the resonances, have very little to do in these asymmetries. It should also be emphasized that the procedure of kinematically ``cutting'' out the resonances can be done both in theory and experiments leaving equivalent impacts on the observables.

\bibliographystyle{jhep}
\bibliography{RS_RC4}

\end{document}